%% file: paper.tex
\documentclass[fleqn,usenatbib]{mnras}

\usepackage{newtxtext,newtxmath}

\usepackage[T1]{fontenc}
\usepackage{ae,aecompl}

\usepackage{graphicx}	
\usepackage{amsmath}	

\usepackage{scalerel}
\usepackage{tikz}
\usetikzlibrary{svg.path}

\definecolor{orcidlogocol}{HTML}{A6CE39}
\tikzset{
  orcidlogo/.pic={
    \fill[orcidlogocol] svg{M256,128c0,70.7-57.3,128-128,128C57.3,256,0,198.7,0,128C0,57.3,57.3,0,128,0C198.7,0,256,57.3,256,128z};
    \fill[white] svg{M86.3,186.2H70.9V79.1h15.4v48.4V186.2z}
                 svg{M108.9,79.1h41.6c39.6,0,57,28.3,57,53.6c0,27.5-21.5,53.6-56.8,53.6h-41.8V79.1z M124.3,172.4h24.5c34.9,0,42.9-26.5,42.9-39.7c0-21.5-13.7-39.7-43.7-39.7h-23.7V172.4z}
                 svg{M88.7,56.8c0,5.5-4.5,10.1-10.1,10.1c-5.6,0-10.1-4.6-10.1-10.1c0-5.6,4.5-10.1,10.1-10.1C84.2,46.7,88.7,51.3,88.7,56.8z};
  }
}

\newcommand\orcidicon[1]{\href{https://orcid.org/#1}{\mbox{\scalerel*{
\begin{tikzpicture}[yscale=-1,transform shape]
\pic{orcidlogo};
\end{tikzpicture}
}{|}}}}

\usepackage{hyperref} 

\newcommand{\lya}{Ly$\alpha$}
\newcommand{\HI}{H\,{\sevensize I}}

\usepackage{soul,color}

\expandafter\newcommand\csname J0827\endcsname{1}
\expandafter\newcommand\csname J0859\endcsname{2}
\expandafter\newcommand\csname J0002\endcsname{3}
\expandafter\newcommand\csname J0759\endcsname{4}
\expandafter\newcommand\csname J0854\endcsname{5}
\expandafter\newcommand\csname J1307\endcsname{6}
\expandafter\newcommand\csname J1352\endcsname{7}
\expandafter\newcommand\csname J2305\endcsname{8}
\expandafter\newcommand\csname J0148\endcsname{9}
\expandafter\newcommand\csname J0826\endcsname{10}
\expandafter\newcommand\csname J0923\endcsname{11}
\expandafter\newcommand\csname J1000\endcsname{12}

\newcommand{\cmmnt}[1]{}

\title[The Luminosity Dependence of Giant Ly$\alpha$ Nebulae]{Revealing the Impact of Quasar Luminosity on Giant Ly$\alpha$ Nebulae}

\author[R. Mackenzie et al.]{Ruari Mackenzie$^{1}$\thanks{E-mail: mruari@phys.ethz.ch}\orcidicon{0000-0003-0417-385X},
Gabriele Pezzulli$^{1,2}$,
Sebastiano Cantalupo$^{1}$,
Raffaella A. Marino$^{1}$, \newauthor
Simon Lilly$^{1}$,
Sowgat Muzahid$^{3}$,
Jorryt Matthee$^{1}$,
Joop Schaye$^{4}$,
Lutz Wisotzki$^{3}$
\\
\\
$^{1}$Department of Physics, ETH Zurich, Wolfgang-Pauli-Strasse 27, 8093 Zurich, Switzerland \\
$^{2}$Kapteyn Astronomical Institute, University of Groningen, Landleven 12, 9747 AD Groningen, The Netherlands \\
$^{3}$Leibniz-Institut f$\ddot{u}$r Astrophysik Potsdam (AIP), An der Sternwarte 16, D-14482 Potsdam, Germany \\
$^{4}$Leiden Observatory, Leiden University, PO Box 9513, NL-2300 RA Leiden, the Netherlands }

\date{Accepted 16 October 2020. Received 13 October 2020; in original form 25 June 2020}

\pubyear{2020}

\begin{document}
\label{firstpage}
\pagerange{\pageref{firstpage}--\pageref{lastpage}}
\maketitle

\begin{abstract}
We present the results from a MUSE survey of twelve $z\simeq3.15$ quasars, which were selected to be much fainter ($20<i_{\rm SDSS}<23$) than in previous studies of Giant \lya\ Nebulae around the brightest quasars ($16.6<i_{\rm AB}<18.7$). We detect \HI\ \lya\ nebulae around 100\% of our target quasars, with emission extending to scales of at least 60 physical kpc, and up to 190 pkpc. We explore correlations between properties of the nebulae and their host quasars, with the goal of connecting variations in the properties of the illuminating QSO to the response in nebular emission. We show that the surface brightness profiles of the nebulae are similar to those of nebulae around bright quasars, but with a lower normalization. Our targeted quasars are on average 3.7 magnitudes ($\simeq30$ times) fainter in UV continuum than our bright reference sample, and yet the nebulae around them are only 4.3 times fainter in mean \lya\ surface brightness, measured between 20 and 50 pkpc. We find significant correlations between the surface brightness of the nebula and the luminosity of the quasar in both UV continuum and \lya. The latter can be interpreted as evidence for a substantial contribution from unresolved inner parts of the nebulae to the narrow components seen in the \lya\ lines of some of our faint quasars, possibly from the inner CGM or from the host galaxy's ISM. 
\end{abstract}


\defcitealias{Borisova}{B16}



\section{Introduction}

In the current paradigm of galaxy formation, galaxies acquire fuel for star-formation by accretion from the intergalactic and circumgalactic media (IGM and CGM). Studying the CGM of high redshift galaxies can shed light on key galaxy evolution processes, such as accretion and outflows. Historically the CGM has been primarily probed through absorption line studies, which have excellent kinematic and diagnostic information, but cannot reveal the overall morphology of the medium. Investigating the CGM though emission is thus a complementary probe, but is constrained by the low expected surface brightness of the CGM of typical galaxies, i.e. in the absence of powerful energy sources such as a luminous central ionizing source. Stacking deep narrow-band data showed \lya\ halos around low-mass galaxies at high redshift, consistent with CGM emission (e.g. \citealp{Hayashino2004, Steidel2011}).  Attempts to use fluorescent emission, powered by the UV continuum of quasars, resulted in the detection of Giant \lya\ Nebulae around hyper-luminous quasars, with sizes up to 460 pkpc \citep{Slug,Jackpot}. However, further narrow-band surveys initially suggested that such large nebulae are rare at $z\simeq2$ \citep{FAB2016}.

Huge progress came with advent of high-throughput Integral Field Spectrographs (IFS) on 8-10 m telescopes, such as the Multi-Unit Spectroscopic Explorer (MUSE, \citealp{BaconMUSE}) and the Keck Cosmic Web Imager (KCWI, \citealp{KCWI}). \cite{Borisova} revealed that Giant \lya\ Nebulae are in fact ubiquitous  around hyper-luminous quasars at $z=3-4$, with a detection rate of 100\%.  The integral field spectroscopy enabled blind searches for nebulae, as well as optimal extraction of the emission in 3D and accurate subtraction of the bright quasar PSF. MUSE also identified \lya\ halos around individual star-forming galaxies for the first time \citep{Wisotzki2016, Leclercq2017}. \cite{MUSEUM} extended quasar studies with MUSE to a larger sample of 61 giant nebulae, again targeting $z=3-4$ quasars with luminosities only slightly lower than those of \cite{Borisova}. More targeted efforts have also studied the nebulae of special QSOs, such as Broad Absorption Line objects \citep{Ginolfi2018}, systems with proximate Damped \lya\ absorption (pDLAs, \citealp{North2017,Marino2019}) and $z\simeq6$ QSOs (e.g. \citealp{Drake2019, REQUIEM}). Recently, \cite{Cai2019} completed a survey of nebulae at $z\simeq2.2$ to explore the evolution of the CGM in massive halos. They found that at this lower redshift the nebulae have lower azimuthally-averaged  surface brightness profiles, which appears to be partially due to more asymmetric emission. The authors further showed that previous narrow-band surveys, which had indicated that Giant \lya\ Nebulae were rare, were hampered by the large uncertainties in estimating quasar systemic redshifts from the broad emission lines.

Despite these recent observational successes, the emission mechanism powering Giant \lya\ Nebulae remains unclear. Some authors have proposed recombination radiation as the dominant mechanism, with the strong ionizing continuum of the quasar powering the emission (\citealp{Cantalupo2005, Slug, Kollmeier2010}). Others have suggested that instead the extended emission is due to the resonant scattering of \lya\ photons from the quasar (e.g. \citealp{Slug}). The contribution of collisional excitation is also debated (see e.g. \citealp{Haiman2000,Fardal2001, Dijkstra2006}).  

Different mechanisms are expected to have a different dependence on the ionization state of the emitting gas and therefore on the intensity of the ionizing radiation. For this reason, we have targeted fainter $z\simeq3$ quasars ($20<i_{\rm SDSS}<23$\footnote{The $i_{\rm SDSS}$ band spans observed wavelengths of 7000-8000 \AA, hence $\sim$1700-1950 \AA\ in the restframe at $z=3.15$.}), in contrast to previous searches that have typically only studied the bright end of the population ($i_{AB}<19$). The goal of this program is to study the response of the nebulae to the drastically different illumination of faint QSOs compared to luminous ones, to search for insights into the emission mechanism behind Giant \lya\ Nebulae. Exploring the regime of lower QSO luminosity could also attenuate a major source of ambiguity in modeling quasar nebulae, the uncertainty of the halo masses of the quasars \citep{Pezzulli2019}. As fainter quasars are much more numerous, their clustering, and hence halo mass, has been measured precisely (e.g. \citealp{Font-Ribera2014,Eftekharzadeh2015}), whereas the halo masses of hyper-luminous quasars is still debated. 

Throughout this paper we report magnitudes in the AB system. We adopt a standard $\Lambda$CDM cosmology with cosmological parameters from \cite{Planck2015}, with H$_0$=67.7 km Mpc$^{-1}$ s$^{-1}$, $\Omega_m$=0.31 and $\Omega_\Lambda$=0.69. We draw extensive comparisons to the sample of brighter quasars observed with MUSE from \cite{Borisova} (henceforth, \citetalias{Borisova}). Note that the properties of the \citetalias{Borisova} nebulae shown below have been remeasured from the original datacubes, this is to ensure consistency in methodology between bright and faint quasar samples. All flux, magnitude and surface brightness measurements have been corrected for Milky Way extinction, using the dust maps of \cite{Planckdust}. These E(B-V) values were converted to band and wavelength specific values using the reddening law of \cite{CCM89}, using $R(V)=3.1$.

\setcounter{footnote}{0}
\begin{figure}
	\centering
	\includegraphics[width=\columnwidth]{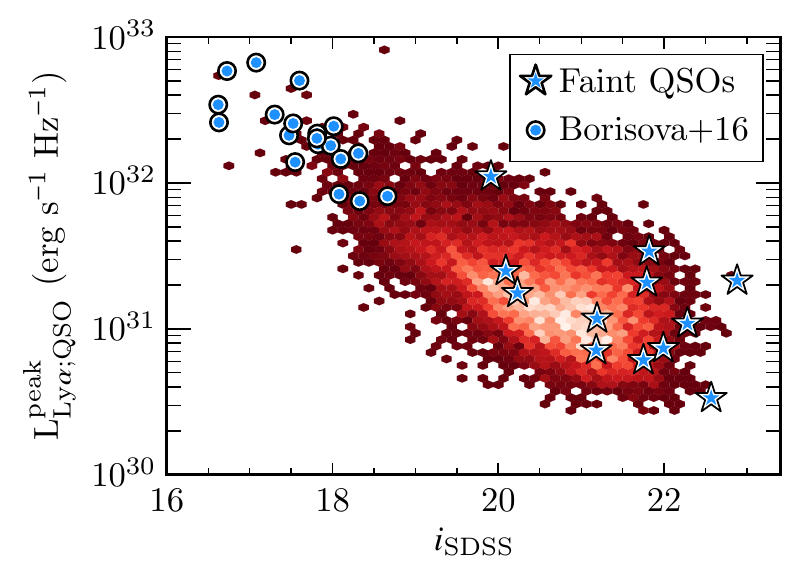}
	\caption{The faint quasar sample in terms of the two parameters used in the selection, the $i_{\rm SDSS}$ apparent magnitude and the peak \lya\ luminosity. For comparison the bright reference sample of \citetalias{Borisova} is also plotted. The parent sample of quasars from SDSS with $3.1<z<3.25$ is also shown for reference, with the density of objects increasing from dark red to pale red.\protect\footnotemark}
	\label{fig:target_selection}
	\vspace{-1em}
\end{figure}

\footnotetext{For consistency, the $i$ band magnitudes shown in Fig. \ref{fig:target_selection} for the MUSE faint sample are the original values from the SDSS database, which were used in selecting the sample \citep{SDSS_DR15}. The photometry reported in Table \ref{tab:sample} is instead derived from the MUSE data. Most of the \citetalias{Borisova} sample fall outside the SDSS survey footprint, so for the \citetalias{Borisova} sample we use synthetic photometry from MUSE in Fig. \ref{fig:target_selection}. In the rest of the paper we will use only the MUSE photometry.}

\section{Observations and analysis}

In this section we will describe the design of the survey, including target selection, observational strategy and data reduction.

\subsection{Sample selection} \label{sec:selection}

\input{tables/tab1_sample_MUSEphot.tex}

We required our targets to have high-quality archival spectra available, to confirm their redshifts and remove systems affected by strong absorption (e.g. broad absorption or proximate Damped \lya\ absorption). In practice this means that our targets are primarily selected from the SDSS and BOSS surveys \citep{SDSS, BOSS}. A single target (J1000+0223, \#\csname J1000\endcsname) is drawn from the zCOSMOS deep survey \citep{zCOSMOSa, zCOSMOSb} in an attempt to probe quasars even fainter than the SDSS selection. Our initial selection was a rough cut at $i_{\rm SDSS}>20$ and a narrow redshift range ($3.1<z<3.25$), where \lya\ falls at wavelengths where the throughput of MUSE is high and there are no sky lines. Within this cut we targeted quasars over a very wide range in apparent magnitude ($i_{\rm SDSS}$) and \lya\ luminosity (measured at the peak of the line), as we believed these two parameters would be relevant to better investigate the recombination and scattering scenarios. Table \ref{tab:sample} lists the properties of the MUSE faint quasar sample. We have ordered the quasars by $i$ band apparent magnitude, with \#1 being the most luminous and \#12 being the faintest. We have remeasured the $i$ band magnitudes for our quasars from the MUSE datacubes, as the original SDSS photometry is outdated and has a low signal-to-noise ratio.

None of our objects are detected in the FIRST radio survey \citep{FIRSTBecker, FIRSTcat}, only $\sim5\%$ of quasars with $i>20$ and $3.1<z<3.25$ in SDSS have matched FIRST detections. Based on the non-detections we can only place an upper-limit on the radio-loudness (i.e. $\rm R=f_{\nu, 5 {\rm GHz}}/f_{\nu, 4400\mathring{A}}$;  \citealp{RadioLoudness}). The median $3\sigma$ limit for our sample is $\rm R<30$, therefore we cannot conclude that all of our objects are radio-quiet ($\rm R<10$), due to their faint nature. However, we expect the majority to be radio-quiet. One source, \#\csname J1000\endcsname, is covered by deeper VLA data in the COSMOS field \citep{VLA_COSMOS} and is not detected. Although this quasar is the faintest within our sample, the deep data show that it is radio quiet ($\rm R<6.7$). \cite{MUSEUM} reported few differences between radio-loud and quiet samples of quasar nebulae, so this parameter may not be significant. 

Fig. \ref{fig:target_selection} compares our sample in terms of $i_{\rm SDSS}$ magnitude and peak \lya\ luminosity against the parent population from SDSS. For the MUSE observed objects the \lya\ luminosity was measured from datacubes directly, whereas for the background population these values are calculated from the SDSS spectra. Combining the fainter quasar sample presented in this paper and the reference sample of \citetalias{Borisova}, we sample 7 magnitudes in UV absolute magnitude (i.e. a factor of $\simeq630$ in luminosity) and a factor of $\simeq$280 in peak \lya\ luminosity. Although our sample size is modest, our dynamic range (in logarithmic space) is more than double that of previous some work which attempted to search for a physical link between nebulae and quasars \citep{MUSEUM}.

\subsection{Observations and data reduction}

In order to make homogeneous comparisons with the bright sample (\citetalias{Borisova}), we followed the same observational strategy. We observed each field for a total time of 3600 s split across 4 exposures, with instrument rotations of 90$^{\circ}$ and dithers of $1''-2''$ between each exposure. Before the first exposure in each field we attempted to offset the quasar from the center of the field, to avoid the poor data quality present in the gaps between stacks of IFUs. Differently from \citetalias{Borisova}, we conducted observations in the WFM-AO-N mode, whereas the bright sample were observed before the commissioning of the adaptive optics system. This difference should have minimal effect, as we are observing at blue wavelengths where the correction is less effective. 

The data were obtained using MUSE on VLT/UT4 using the AOF+GALACSI adaptive optics system \citep{GALACSI} between the 13$^{\rm th}$ of September 2018 and the 3$^{\rm rd}$ of May 2019 in dark time. The spatial full width at half maximum measured in the final datacubes varies between 0.74'' and 1.38'', this is calculated from collapsed images spanning 4800-5300 \AA, centered on the median QSO \lya\ wavelength. Measured over 5000-9000 \AA, the FWHM ranges from 0.73'' to 1.13''.

The initial data reduction was carried out using the MUSE pipeline (v2.6, \citealp{MUSEPipeline,MUSEPipeline2}), which applied bias subtraction, flat-fielding, wavelength and astrometric calibration, and flux calibration using a standard star observation. The quality of the basic reduction is limited by the flat-fielding, so we perform additional self-calibration post-processing to improve the illumination correction with CubEx (Cantalupo in prep. also see e.g. \citetalias{Borisova}, \citealt{Marino2018,Marino2019,Cantalupo2019}). Due to the the low number of exposures per field (4) we combined the exposures using median statistics. This was done because the reduction software cannot adequately reject cosmic rays with such few exposures. \citetalias{Borisova} also used median coadded data for this reason. Lastly we checked the spectrophotometric calibration of the datacubes with aperture photometry of stars in synthetic $r$ and $i$ band images against SDSS. None of our datacubes required re-calibration. 

\subsection{QSO PSF and continuum subtraction}

Before searching for extended emission, we first removed the continuum light from the quasars, using CubePSFSub (a utility within CubEx). First we subtracted the quasar continuum emission using an empirical PSF estimation. The routine estimates the PSF using the quasar continuum by constructing a narrow-band image around each layer, composed of 300 channels. The image of the QSO is used as the empirical PSF for that layer. The width of the psudo-narrow-band image is larger than the 150 channels used in \citetalias{Borisova}, as our fainter quasars required more channels to reduce the noise in the PSF estimation. Negative pixels in the PSF image which are more than $1\sigma$ below zero flux are clipped and set to zero, to prevent negative noise spikes in the PSF image that would artificially make the nebulae brighter. The constructed PSF image is then scaled to match the flux in a $1''\times1''$ area centered on the QSO and is re-centered on the peak emission in each layer. This scaled and centered empirical PSF is then subtracted from a given layer, up to maximum radius of $3''$. In \citetalias{Borisova} this radius was $5''$, but our much fainter quasars are not bright enough to have strong PSF wings. During this procedure the channels around the quasar \lya\ peak are masked in the construction of the narrow-band images (30 to 50 layers), which prevents the extended emission from being present in these PSF images and being subtracted. 

The PSF-subtracted cubes still contain continuum light from sources other than the quasar. As these are typically much fainter than the quasars and can have a range of morphologies, we subtract their continuum emission by median filtering the cube at each spatial location. The cube is binned in the spectral direction, with a size of 80 to 150 pixels using median statistics, then the spectra are smoothed with a filter size of two bins. As before, the layers around the quasar \lya\ line are masked and continuum subtraction is interpolated across this range. This filtered cube is then subtracted from the cube from the previous step. 

Finally, the cubes were trimmed in wavelength, centered on the QSO \lya\ line plus a margin of $\simeq$200 channels on either side.  

\subsection{Nebula detection}\label{sec:neb_detect}

\begin{figure*}
	\includegraphics[width=0.98\textwidth]{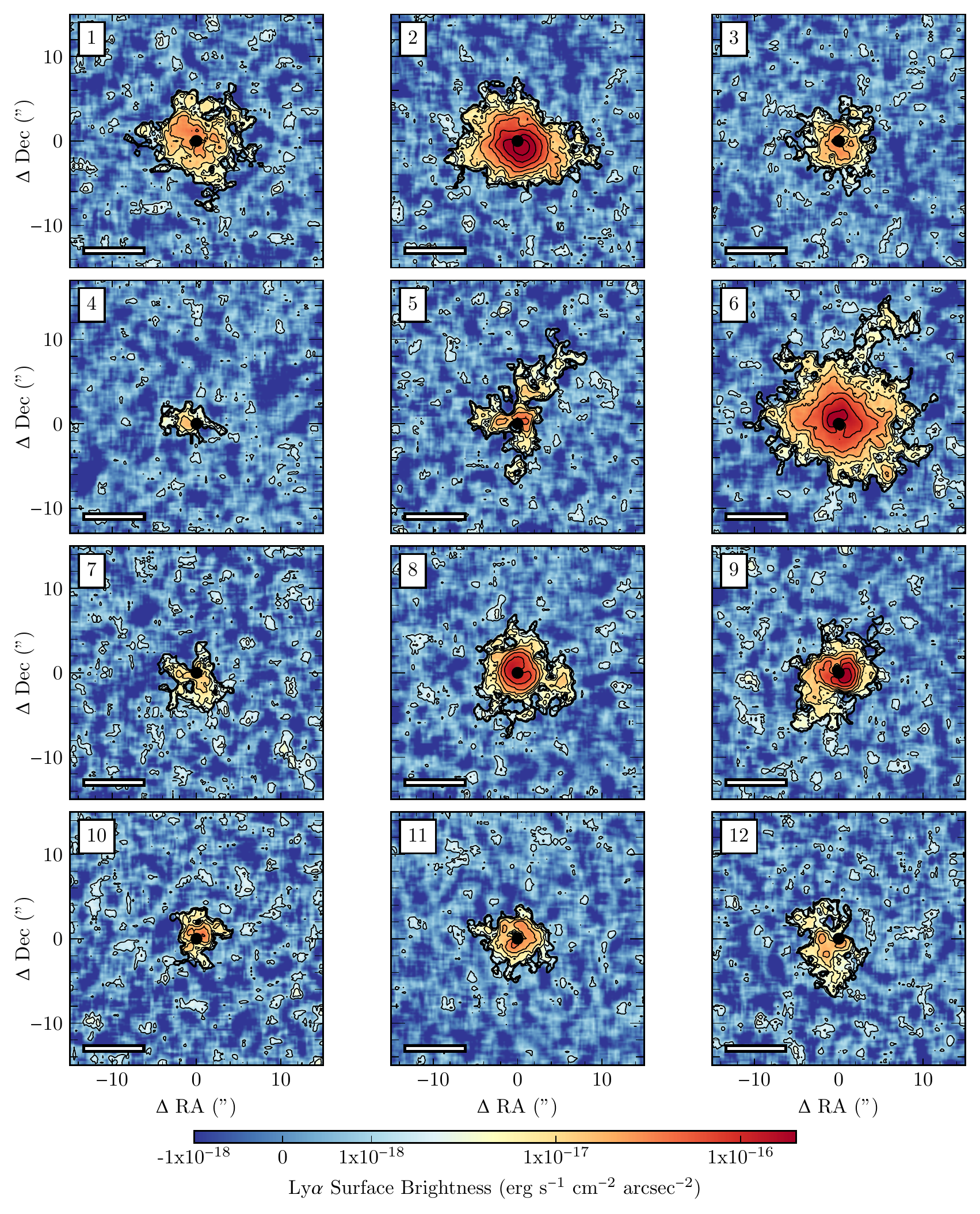}
	\caption{Optimally extracted \lya\ surface brightness maps of our newly discovered nebulae around faint QSOs at $z\simeq3$. Each image is shown on the same angular scale (30x30 arcsec), with the coordinates in arcseconds offset from the QSO, indicated by the black circle. Prior to extraction, the datacubes were filtered with a 3x3 pixel (0.6x0.6 arcsec) boxcar in the spatial directions. Thin contours indicate surface-brightness levels in the image, where each contour is double the last ($1.25\times10^{-18}$, $2.5\times10^{-18}$, $5\times10^{-18}$... $3.2\times10^{-16}$ erg s$^{-1}$ cm$^{-2}$ arcsec$^{-2}$), while the thick black contour indicates the extent of the 3D detection of the nebula (out to SNR=2 per voxel). The white bar in each image indicates 50 physical kpc at the redshift of the quasar. Images are typically centered on the QSO, but those in the second row are shifted by 2'' vertically to accommodate the extent of nebula 6.}
	\label{fig:SB_maps}
	\vspace{-1em}
\end{figure*}

With the contaminating light of continuum sources and the QSO PSF removed, the nebulae can be extracted. For this task we again use CubEx, which performs source detection of objects in 3-dimensional data. First the variance of each cube is rescaled in each layer to match the standard deviation of the flux values in the cube. This is necessary as the propagated variance is typically underestimated due to the resampling of datacubes in the basic data reduction process. The cube is then filtered (smoothed) in the spatial directions, using a Gaussian kernel with a radius of 2 pixels. Next voxels above a signal-to-noise threshold of 2 are grouped together into contiguous groups. We use a low signal-to-noise ratio (SNR) threshold to try to capture the full extent of the nebulae. Groups below a specified minimum number of voxels are removed from the list. Initially this minimum is 4000 voxels, but if this is too high to detect the nebula it is lowered. For our sample the lowest value of this parameter was 1000 voxels. The detection thresholds used follow those of  \citetalias{Borisova}. The voxels identified as part of the nebula group become the 3D segmentation map or mask, which is used to define the extent and properties of each nebula.

\section{Results}\label{sec:Results}

As a result of the detection procedure described in the previous section, nebulae are detected in all 12 fields. Even for quasars up to 6 magnitudes fainter than those of \citetalias{Borisova}, MUSE is still able to recover nebulae in just 1 hour of integration time with a 100\% success rate. The measured properties of the nebulae are summarised in Table \ref{tab:nebulae}. 

In this section we discuss the observed properties of the faint QSO \lya\ nebulae and compare them to the nebulae around luminous QSOs. Following \citetalias{Borisova}, we will discuss spatial and spectral morphology and the surface brightness profiles of the nebulae. However, we do not attempt to extract 2D kinematic maps, as our sample spans a large range in SNR. 

\input{tables/tab2_measurements_MUSEphot}

\subsection{Spatial morphology}\label{sec:morph}

Fig. \ref{fig:SB_maps} shows the optimally extracted surface-brightness maps of our newly discovered \lya\ nebulae. Optimal extraction utilizes a 3D-mask, as described in Sect. \ref{sec:neb_detect}, which encloses the voxels identified as part of the nebula. All voxels within the mask are projected onto a 2D image, and a single layer from outside the mask is added to give some impression of the noise level. See \citetalias{Borisova} for further details. To aid in the comparison all the images are shown on the same angular scale. 

It is noticeable how small these nebulae are in comparison to those of \citetalias{Borisova}. The maximum extent is defined by the 3D-mask described previously, as the maximum projected distance across the mask. \citetalias{Borisova} found that each QSO nebula had an extent of $>100$ pkpc. However for the fainter QSOs, Table \ref{tab:nebulae} shows that only half of the fainter sample extend to this size. While the median size in the reference bright QSO sample is 180 pkpc\footnote{Note that for consistency we have remeasured the sizes of the \citetalias{Borisova} nebulae. As we use a different masking scheme some of the nebulae have larger sizes than reported in the original paper, although most are very similar. The new size estimates are included in Table \ref{tab:nebulae_borisova} for completeness.}, for the fainter QSO sample it is only 100 pkpc. For a visual impression of the scale of this difference, Nebula \#\csname J1307\endcsname\ is fairly average in size in comparison to the nebulae of \citetalias{Borisova}. Fig. \ref{fig:SB_maps} shows that most nebulae in the faint QSO sample are much less extended and luminous than Nebula \#\csname J1307\endcsname, and hence, than a typical bright QSO nebula. Note however that, as the size is defined here by a SNR (and therefore SB) threshold, smaller sizes can be explained as a consequence of the nebulae being fainter (see Sect. \ref{sec:SBprofs}). We do not detect any very extended nebulae ($>200$ pkpc), of which 2 examples were found in the \citetalias{Borisova} sample of 19 QSOs. With our modest sample size, however, we cannot exclude a similar occurrence rate among low-luminosity QSOs. There is also no obvious evidence of a correlation between size and asymmetry, as was suggested in \citetalias{Borisova}. 

Half of the detected nebulae appear to be close to circularly symmetric. In the inner regions a few nebulae are brighter on one side of the QSO than the other (e.g. \#\csname J0859\endcsname\ and \csname J0148\endcsname)\cmmnt{\#3 and 9)}, this is also similar to some cases in the \citetalias{Borisova} sample. For some of the faintest surface brightness nebulae, the emission is only detected on one side of the QSO (\#\csname J0759\endcsname\ and \csname J1000\endcsname)\cmmnt{(\#10 and 12)}, but we cannot exclude the possibility that these are analogous to the previous case of asymmetry but rescaled to lower surface brightness and truncated by the SNR limit. Two objects show evidence of filamentary structures (\#\csname J0854\endcsname\ and \csname J1307\endcsname)\cmmnt{(\#4 and 6)}, although much less extended than the spectacular examples of \citetalias{Borisova}. \#\csname J0854\endcsname\ and \csname J1307\endcsname\ are however quite different from each other. \#\csname J1307\endcsname\ is dominated by a circularly symmetric component with the filament just barely above the detection threshold. This type of asymmetry could be common but it might simply fall below the detection limit in fainter nebulae. \#\csname J0854\endcsname\ seems to have an intrinsically asymmetric morphology. In Appendix \ref{ap:asym} we have attempted to quantify the asymmetry of the quasar nebulae using a technique adopted in previous studies of Ly$\rm \alpha$ nebulae. On average the faint quasar nebulae are more circularly symmetric than those around bright quasars, but this difference is not statistically significant. Overall, the morphologies of our fainter QSO nebulae are similar to those of \citetalias{Borisova}, even though the quasars are much less luminous.

\subsection{Spectral properties}\label{sec:spectra}

\begin{figure*}
	\centering
	\includegraphics[width=\textwidth]{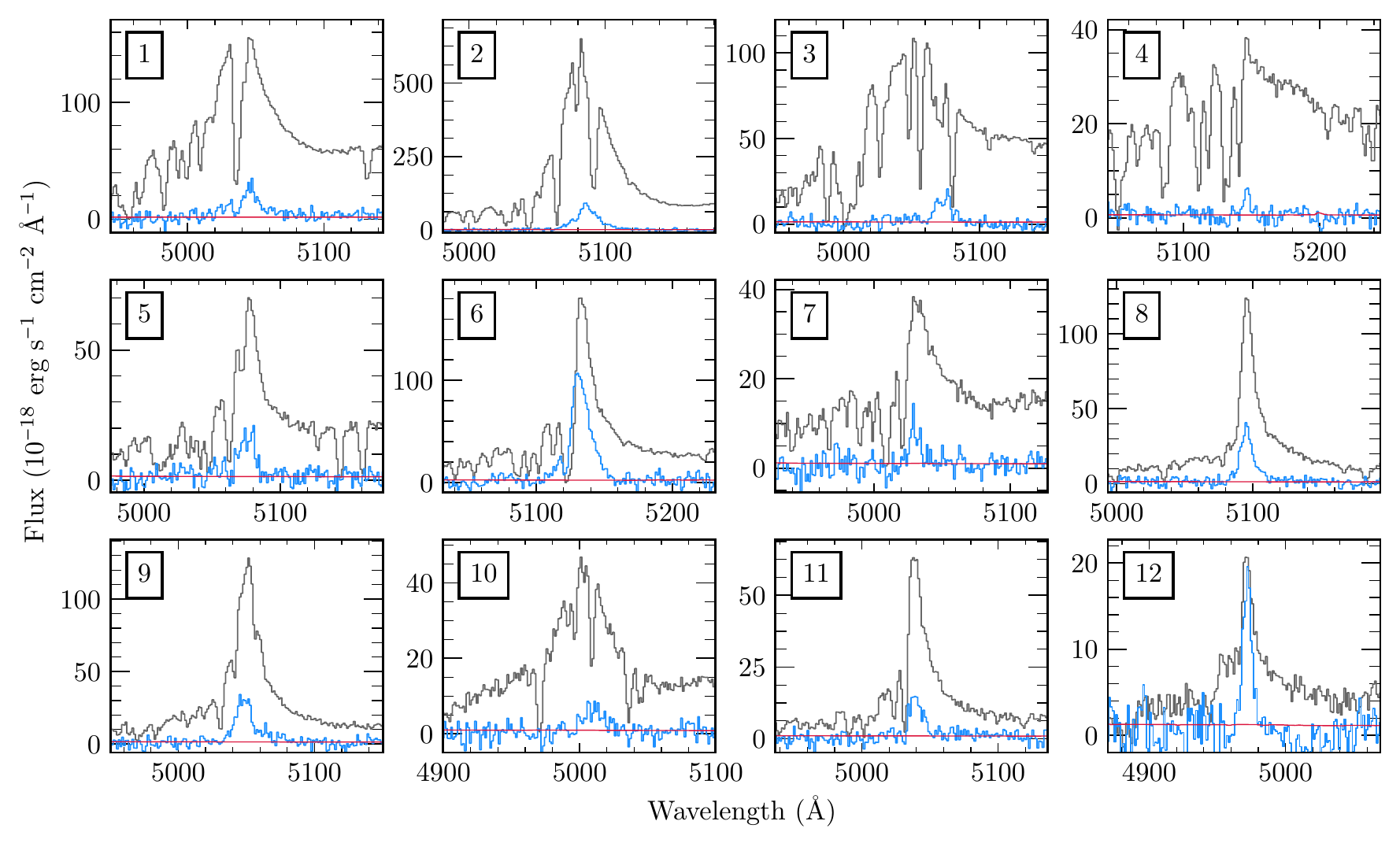}
	\caption{Spectra of the \lya\ line of each nebula (blue line), shown for context with the spectrum of their QSO (grey line). The 1$\sigma$ noise of the nebula spectrum is indicated (red). The nebula spectra are extracted by first projecting the 3D mask onto 2D and then using that projected mask to extract the spectrum from the continuum subtracted cube. The spectrum of each QSO was extracted from the unsubtracted MUSE cube using a 2'' radius aperture. Both spectra are shown on the same scale.}
	\label{fig:spectra}
	\vspace{-1em}
\end{figure*}

Fig. \ref{fig:spectra} shows the extracted 1D spectral profile of the \lya\ line of each nebula, and compares it with that of the host QSO. All spectra are extracted from the MUSE datacubes for consistency. In each case the nebula spectrum is taken by projecting the 3D-mask into a 2D-mask (x,y) and summing the spectra at every spatial location in the mask. The extent of the 2D-mask is shown by the thick contour in Fig. \ref{fig:SB_maps}. The QSO position is masked within a radius of 2.5 pixels when extracting the nebula spectrum, as there are frequently large residuals from the PSF subtraction. The cube used to extract the nebula spectrum is the detection cube with QSO PSF and continuum light subtracted. The QSO spectrum is extracted using an aperture with a radius of 2'' on the original (unsubtracted) datacube.

\newpage When compared to the brighter QSOs of \citetalias{Borisova}, the fainter QSOs often display \lya\ lines dominated by a narrow core. QSO \#\csname J1307\endcsname\ is perhaps the most extreme example. The lines do appear to have broad \lya\ wings however, and the QSOs have other broad lines. Hence, these are still classified as broad-line QSOs. Our sample also includes QSOs with \lya\ line morphologies more typical of bright QSOs, such as QSOs \#\csname J0002\endcsname\ and \csname J0759\endcsname\cmmnt{\# 2 and 10}. The increasing equivalent width of quasar emission lines with decreasing UV luminosity is known as the Baldwin effect \citep{Baldwin1978,baldwin_REV} and it is dominated by a narrow component of the lines \citep{Osmer1994}. When considering the spectra of the nebulae of the QSOs with strongly peaked \lya\ emission, the line shapes appear to bear striking resemblance to the narrow peaks of the QSO lines (e.g. \#\csname J2305\endcsname, \csname J1307\endcsname, \csname J0923\endcsname\ and \csname J0148\endcsname)\cmmnt{(e.g. \#1, 3, 5 and 6)}. One very clear example is \#\csname J1000\endcsname\cmmnt{\#12}, where the QSO \lya\ line has a narrow peak on top of a much broader component. In this case, the \lya\ line of the nebula very clearly matches the narrow component of the QSO profile. In this extreme example the peak flux of the spectrum of the nebula is even comparable to that of the QSO. It is however important to recall that the aperture used to extract the nebula spectrum is much larger. In Section \ref{subsec:CentralEmission} we discuss a possible explanation of this interesting effect. 

For almost all of our nebulae, the flux centroid of the nebular \lya\ spectrum is very close to the peak of the host QSO's \lya\ line. This is consistent with \cite{MUSEUM} and \cite{OSullivan2019}, who showed quantitatively that the peak of the QSO \lya\ spectrum is a better predictor of the redshift of the \lya\ nebula compared to other estimates of the systemic redshift. One notable exception in our sample is Nebula \#\csname J0002\endcsname\, which is redshifted with respect to the peak of the QSO's \lya\ line by 1360 km/s. Note that the \lya\ line of QSO \#\csname J0002\endcsname\ is clearly affected by multiple absorption lines, which may be somehow related to this offset.

A number of nebulae show signs of absorption lines which coincide with absorption seen in their QSO. \#\csname J0827\endcsname\ and \csname J1307\endcsname\cmmnt{\#6 and 11} are clear examples where the nebular emission is present on both sides of the absorption line. This can be explained in a scenario where the \HI\ absorption system is in front of the QSO and nebula, with the neutral cloud covering a significant (flux-weighted) fraction of the nebula. Coherent absorption across scales of over 100 kpc has been observed in nebulae surrounding high-redshift radio galaxies (e.g. \citealp{Swinbank2015}). Nebula  \#\csname J0859\endcsname\cmmnt{\#9} seems to show the opposite behaviour, where the QSO has an absorption feature just bluewards of the \lya\ peak and yet the nebula shows no coincident absorption. Together, these examples indicate that the \lya\ absorption systems have a wide range of scales, and potentially a variety of distances from the quasar nucleus. We also recall however that not every dip in a \lya\ spectrum is necessarily due to an intervening absorber, as both complex kinematics (e.g. multiple components along the line of sight) and radiative transfer effects can also produce a similar phenomenology in some cases.

\subsection{Surface brightness profiles}\label{sec:SBprofs}

\begin{figure*}
	\centering
	\includegraphics[width=\textwidth]{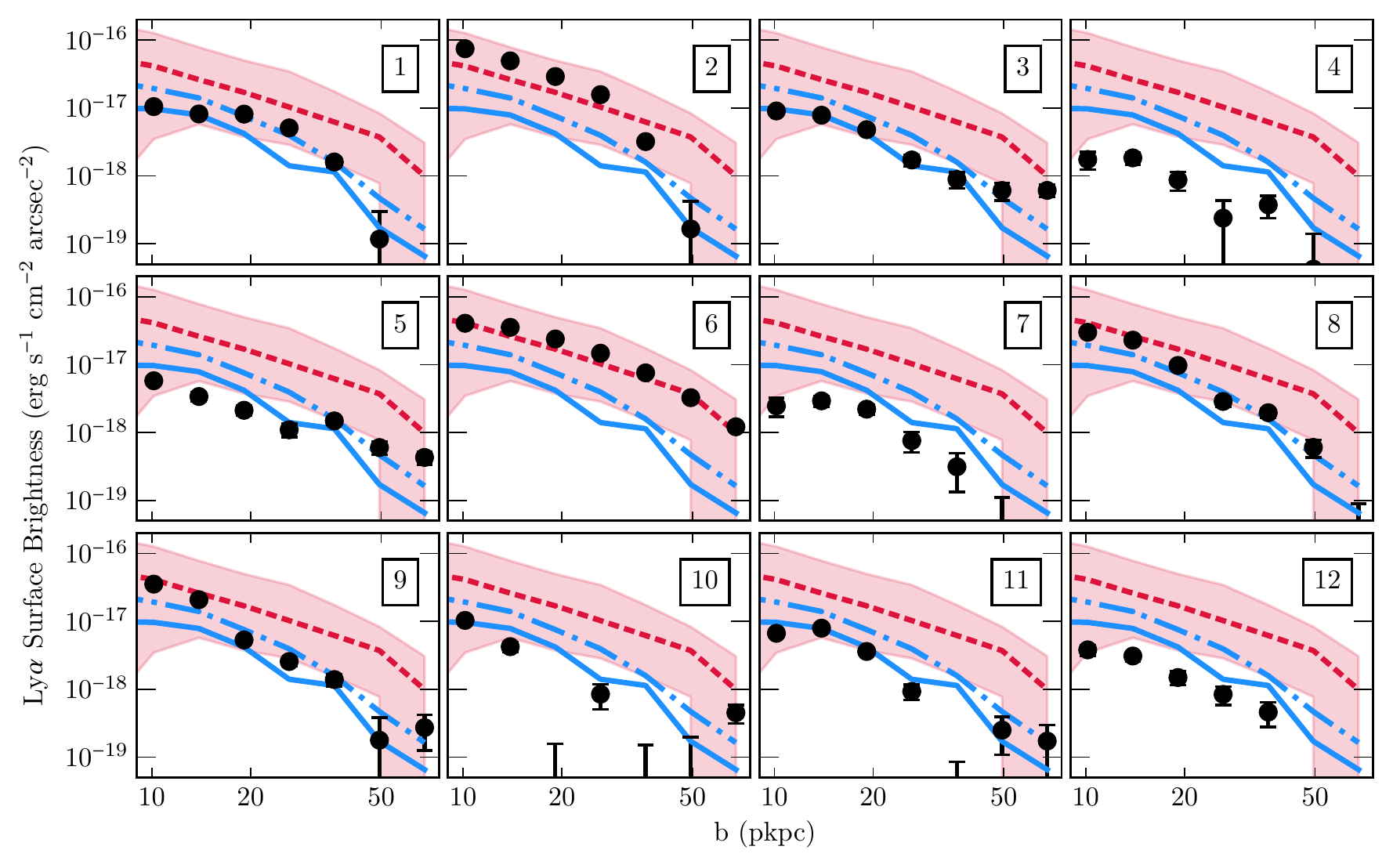}
	\caption{\lya\ surface brightness profiles of the nebulae around the faint QSO sample. Each panel shows the measurements for a given nebula with 1$\sigma$ error bars (black points), along with the mean (blue dot-dashed line) and median (blue solid line) of the faint sample. Both statistics are shown as our sample spans a large range of behaviours. Also plotted for reference is the mean surface brightness profile of the bright QSO sample from \citetalias{Borisova} (red dashed line), and the full range of profiles within that sample (red shaded region). Note that here we have not corrected for redshift dimming, but the samples have only a small spread in redshift.}
	\label{fig:SB_panel}
	\vspace{-1em}
\end{figure*}

As already noted, the \lya\ nebulae around our fainter QSO sample are smaller than those of bright QSOs, however the reported sizes are determined by our observational limits. To better understand the difference, we now extract surface brightness profiles for each of our QSO nebulae. 

The optimally extracted images shown in Fig. \ref{fig:SB_maps} are excellent for capturing the extent and morphology of the nebulae, but the noise properties of these images are complicated by the use of the 3D mask, which by definition applies a SNR threshold \citep{Borisova,MUSEUM}. Following \citetalias{Borisova}, for the purpose of determining surface brightness profiles we use pseudo-narrow-band images extracted from the continuum-subtracted datacubes. The width of each narrow-band is set by the maximum extent of the detected nebula in the 3D mask in the spectral direction. The maximum extent is used to try to capture the full flux of the nebula, although in the outer regions the large number of layers will limit the SNR. We also mask regions in datacubes that are believed to be affected by systematics or the residuals of bright continuum objects. The individual profiles are shown in Fig. \ref{fig:SB_panel}, along with the mean and median of our sample and that of \citetalias{Borisova}. The errors on the surface brightness profiles are propagated from the flux errors in the pseudo-narrow-band images; they do not include uncertainties due to the background or PSF subtractions. All of the profiles and surface brightness measurements for the \citetalias{Borisova} sample have re-estimated from the original data, in order to guarantee uniform analysis across the two datasets (see Appendix \ref{ap:A} for further details and Table \ref{tab:nebulae_borisova} for the remeasured quantities). 

The mean profile of our sample reveal that the fainter QSOs host on-average fainter nebulae than the bright sample (\citetalias{Borisova}). As our sample was selected only on QSO properties, we can therefore conclude that there is a connection between nebula surface brightness and QSO magnitude. Although the faint QSO profiles are on average lower in surface brightness, the mean and median profiles have very similar slopes to that of the bright QSO sample. Looking beyond the typical behavior, it is apparent that there is substantial diversity among the faint QSO sample. Some objects (Nebulae  \#\csname J0859\endcsname\ and \csname J1307\endcsname\cmmnt{\#6 and 9}) are perfectly compatible with the mean profile of the brighter QSO nebulae. QSO \#\csname J0859\endcsname\ has the brightest absolute magnitude in the faint sample, see Table \ref{tab:sample}. On the other hand, QSO \#\csname J1307\endcsname\ has a fainter UV continuum (M$_i (z=2)\simeq$-26), but has one of the brightest QSO \lya\ lines. Even for the much fainter nebulae (e.g \#\csname J0759\endcsname\ and \csname J1000\endcsname\cmmnt{\#10 and 12}), there is still no evidence of a departure from the shape of the mean profile of \citetalias{Borisova}, even with QSOs which are almost 5 magnitudes fainter in UV continuum. In the next section we will leverage this extreme dynamic range to try to understand the link between QSO and nebula properties.

Fig. \ref{fig:SB_comp} (top) shows the surface brightness profiles of all our nebulae, corrected for cosmological dimming to a common $z=3$ (using a factor of $[(1+z)/4]^4$). The mean and median profiles are again shown, along with the mean of the brighter sample. Comparing the profiles to typical surface brightness profiles of \lya\ Emitter (LAE) galaxies \citep{Wisotzki2016}, reveals that the fainter QSO nebulae are still brighter and with flatter surface brightness profiles than the \lya\ halos around normal star-forming galaxies. It can also be seen that there is a larger diversity of surface brightness profiles in the fainter sample than in the bright one. Although the shapes of the profiles are mostly consistent, their normalization in surface brightness seems to vary by over an order of magnitude. 

The comparison between the two samples suggests a trend with UV continuum. To better investigate this trend, we now combine in one single analysis the individual SB profiles of all the nebulae extracted from both samples. This combination is made possible by the consistent observational strategy and data reduction used by this work and \citetalias{Borisova}. Fig. \ref{fig:SB_comp} (lower-left) shows all the surface brightness profiles in the combined sample, where each line has been coloured by the $i$ band absolute magnitude of the QSO (K-corrected to $z=2$, \citealp{Ross2013}). This colour-coding reveals that there is indeed a gradient in UV magnitude going from the brightest to the faintest profiles, seen as a gradient from red to blue with decreasing surface brightness. However, there is clearly a large scatter in this relation as there are UV-faint QSOs among the brightest profiles (e.g. QSO \#\csname J1307\endcsname\cmmnt{6}, as noted above). Fig. \ref{fig:SB_comp} (lower-right) is similar, but now the profiles are labeled by the specific luminosity of the peak of the QSO \lya\ line. This parameter, $\rm L^{peak}_{Ly\alpha;QSO}$, was chosen following \cite{MUSEUM}. Here it is estimated from the peak flux density of the \lya\ line in a MUSE spectrum extracted within a 2'' radius aperture. The peak flux density is then converted to a luminosity density in order to remove the redshift dependence. Fig. \ref{fig:SB_comp} (lower-right) shows a very clear gradient with $\rm L^{peak}_{Ly\alpha;QSO}$ and surface brightness in the range of 20 to 50 pkpc, evidence of a tight correlation between profile normalization and \lya\ peak luminosity density. Looking at smaller radii (10-20 pkpc), the trend is less monotonic than at larger radii (20-50 pkpc). This could be due to the PSF subtraction, which is more uncertain at smaller radii. To assess the strength of this correlation we require a parameter to best capture this result. In the following section we introduce SB$_{\rm outer}$ as the redshift-dimming-corrected \lya\ surface brightness between 20 and 50 pkpc. 

\subsection{Luminosity dependence of \texorpdfstring{Ly$\alpha$}{Lya} nebulae}\label{subsec:lumdep}

\begin{figure*}
	\centering
	\includegraphics[width=0.8\textwidth]{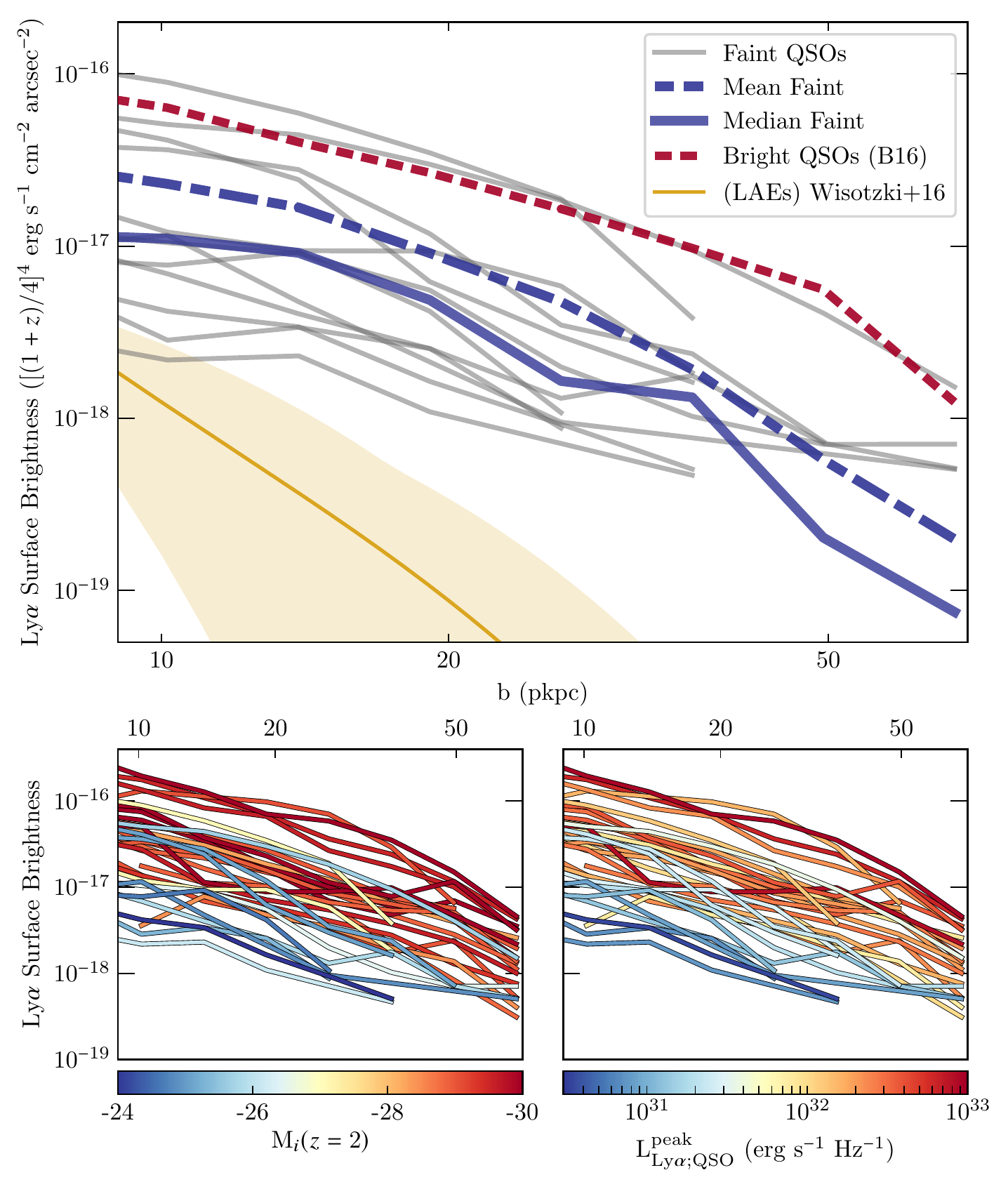}
	\caption{{\bf (Top)} \lya\ surface brightness profiles of the nebulae around the faint QSO sample, now corrected for redshift dimming. The profiles are only plotted where $\rm SNR > 2$. The mean (blue dot-dashed line) and median (blue solid line) profiles of the fainter sample are shown. Also plotted for reference is the mean surface brightness profile of the bright QSO sample from \citetalias{Borisova} (red dotted line). The yellow region indicates typical profiles of haloes around LAEs (so-called \lya\ haloes). {\bf (Lower Left)} The combined profiles of the fainter QSO and \citetalias{Borisova} samples as before, but with the profiles coloured by the $i$ band absolute magnitude of the QSO. K-corrections have been applied to the absolute magnitudes, normalized to $z=2$ (see \citealp{Ross2013}). {\bf (Lower Right)} Similar to lower-left, but now colour-coding the QSO peak \lya\ luminosity density.}
	\label{fig:SB_comp}
	\vspace{-1em}
\end{figure*}

\begin{figure*}
	\centering
	\includegraphics[width=0.7\textwidth]{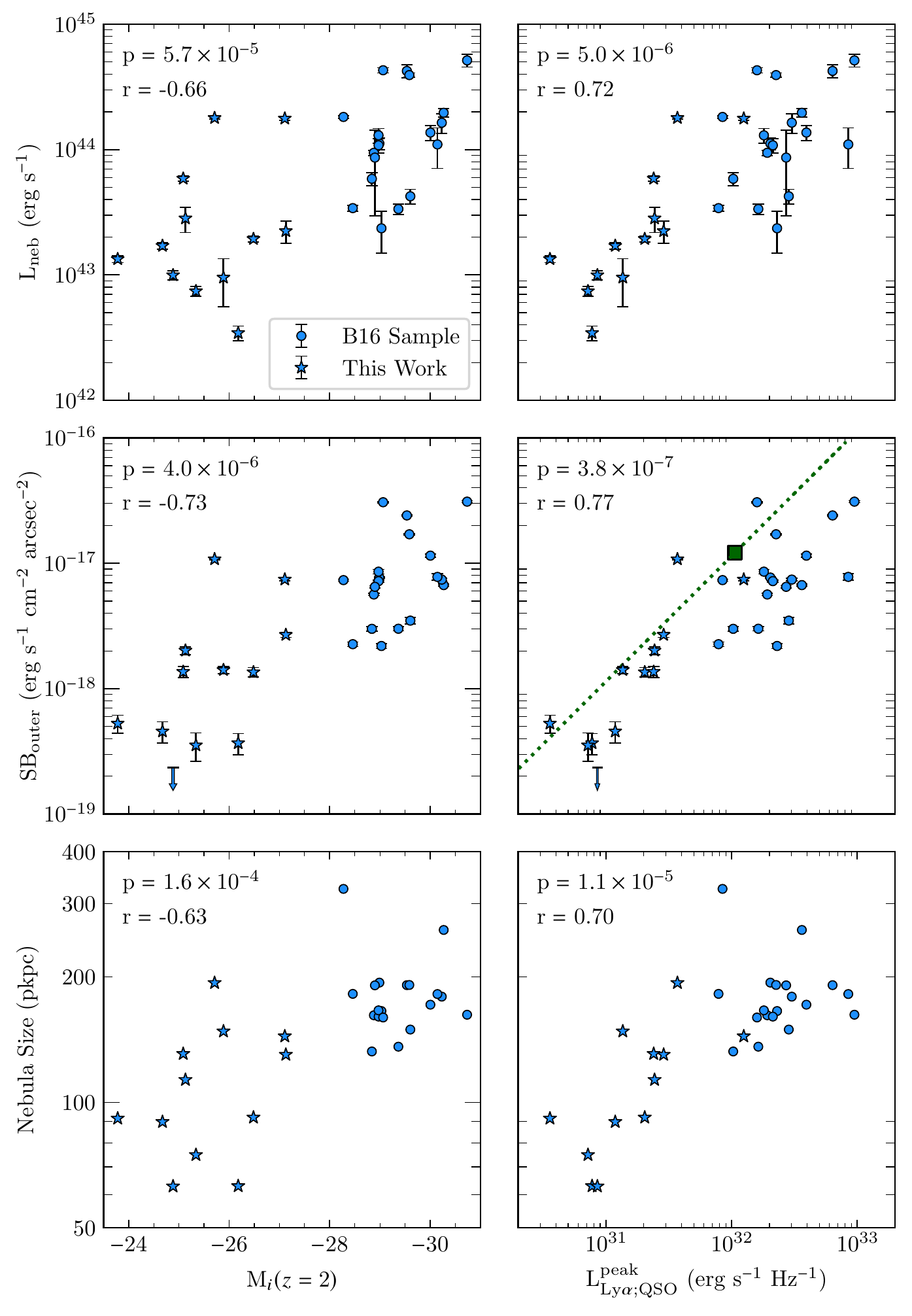}
	\caption{Comparisons between QSO and nebular properties for both the fainter sample (blue stars) and the bright sample (blue circles, \protect\citetalias{Borisova}). The three nebular parameters selected are the total \lya\ luminosity (top row), the (redshift-corrected) surface brightness measured between 20 and 50 pkpc (middle row), and the maximum projected size (bottom row). These properties are plotted against both the QSO $i$ band absolute magnitude (K-corrected to $z=2$; left column) and the peak luminosity density of the QSO \lya\ line (right column). In each panel the significance of the correlation is listed with the Spearman's correlation coefficient r, while the p-values indicate the significance at which the null hypotheses of no correlation can be rejected using this test. In the middle right panel the green square indicates the exotic quasar from \protect\cite{Marino2019}, and the green dashed line is the linear relation which passes through this point. }
	\label{fig:correlations}
	\vspace{-1em}
\end{figure*}

Prompted by the correlations observed in  Fig. \ref{fig:SB_comp}, we now move to a more quantitative analysis of these potential relationships. The panels of Fig. \ref{fig:correlations} show three measured properties of \lya\ nebulae (total \lya\ luminosity, surface brightness between 20 and 50 pkpc, and size, defined as the maximum projected distance across the 3D detection mask) against two QSO properties ($i$ band absolute magnitude and peak \lya\ luminosity density) for the fainter and bright samples. It is important to bear in mind that these nebular properties are not independent of each other. Naturally, \lya\ luminosity and outer surface brightness are correlated properties. Size is also correlated with the other two properties, as the size is defined by the maximum extent above a limiting surface brightness. This analysis follows \cite{MUSEUM}, but we have selected different nebula properties as motivated by the results from the previous section. The total luminosity should be dominated by the bright inner region of the nebula, so it is complemented by the outer surface brightness (SB$_{\rm outer}$). In this section we will emphasize the most remarkable facts emerging from Fig. \ref{fig:correlations}. In the next section (\ref{sec::Discussion}) we will discuss the implications of these results.

The left hand panels of Fig. \ref{fig:correlations} show the nebular properties as a function of $i$ band absolute magnitude, which traces the underlying UV continuum of the QSO. It is shown that fainter QSOs have on average smaller, less-luminous nebulae, with lower outer surface brightness. All three of these correlations are significant, with p-values ranging between $1.6\times10^{-5}$ and $4\times10^{-6}$, as established from a Spearman's rank correlation test. While the correlations are significant, there is also a large scatter. In each property the median is different between the bright and fainter populations, but the range within each of the two samples appears to be larger than these differences. While the nebula properties clearly depend on the QSO's absolute magnitude, the correlations are not tight. Furthermore, the scatter among the faint population seems to be significantly larger than for the bright end, even after accounting for uncertainties. It can be seen that the correlations appear to be shallower than a linear relation. Some previous searches for luminosity dependence, such as \cite{MUSEUM}, lacked the necessary dynamic range in UV magnitude to uncover these relations. \cite{OSullivan2019} and \cite{Farina2019} had samples which, in principle, spanned sufficient range to detect these correlations, but nebulae were not detected for most of the fainter quasars. In both works only two nebulae were detected around M$_i$>-26 quasars. Fig. \ref{fig:correlations} (middle-left) shows that our faint-quasar nebulae span a large range in surface brightness. If one could only detect only the two highest surface brightness nebulae among the faint quasars, then the detections would be consistent with the bright quasar sample, and hence it is possible that some previous findings were biased by this selection effect. Other studies of fainter QSOs typically suffered from low sample size and sensitivity (e.g. \citealp{Christensen2006,Fathivavsari2016}). Our study combines both a large dynamic range and high sensitivity observations, such that our results are not complicated by non-detections. 

As indicated both by Fig. \ref{fig:SB_comp} (bottom-right) and by Fig. \ref{fig:correlations} (middle-right), the correlation between outer surface brightness and peak QSO \lya\ luminosity density is much tighter than that with M$_i(z=2)$, with a corresponding p-value of $\simeq4\times10^{-7}$. Fig. \ref{fig:correlations} reveals that the correlations with the nebula's luminosity and size are also significantly stronger when plotted against peak QSO \lya\ luminosity (right column), rather than $i$ band absolute magnitude (left column). This reduction in the scatter is wholly driven by the fainter sample, as the dispersion among the bright \citetalias{Borisova} sample appears to be unchanged if the nebular properties are plotted against QSO UV luminosity or peak \lya\ luminosity density. The relation between nebula luminosity and $\rm L^{peak}_{Ly\alpha;QSO}$ is significant, although there is considerable scatter. The correlation with outer surface brightness among the fainter sample is striking, but for the bright sample the scatter is much larger. A fit to both samples would be shallower than the linear relation shown. Lastly there is a clear correspondence between peak \lya\ luminosity and size. However, it is not clear whether this relation is fundamental or a consequence of the link between maximum extent and outer surface brightness. 

\section{Discussion}\label{sec::Discussion}

In the results presented above we have clearly established that the luminosity, surface brightness and apparent size of Giant \lya\ Nebulae are correlated with the luminosity of their host QSOs, both in UV continuum and peak \lya\ luminosity. In this section we discuss possible scenarios that could explain these relations. This discussion is structured in the following way: First, we examine if our fainter QSO sample can shed new light on the physical processes responsible for powering the emission in Giant \lya\ Nebulae. Next, we consider if secondary properties may be responsible for the observed luminosity dependence. We then discuss possible explanations for the linear correlation between nebula surface brightness and QSO peak \lya\ luminosity. Finally, we will outline future directions extending beyond this work.

\subsection{Powering mechanism}\label{subsec:PoweringMechanism}

In general, three physical mechanisms can contribute to the Ly$\alpha$ emission from the circumgalactic medium: recombination (in the context of QSO nebulae, this follows photoionization from the quasar UV radiation and is also often called   \emph{ fluorescence}), scattering of Ly$\alpha$ photons originally produced by a central source (also called \emph{photon pumping}; in the case of QSO nebulae, the central source would be the broad line region, or BLR) and collisional excitation (see e.g.\ \citetalias{Borisova} and references therein for a discussion). Whether any of these mechanisms provide a non-negligible, or even dominant, contribution to the total emission depends rather strongly on the physical properties of the emitting medium, including its density, temperature and, above all, its ionization state.

Even relatively faint QSOs are, in absolute terms, very strong sources of ionizing photons. The size of the Str\"omgren sphere associated to the QSOs in our sample can be estimated as $R_\textrm{ion} = 1 \; \dot{N}_\textrm{ion,56}^{1/3} \; \Delta_{200}^{-2/3} \; \textrm{Mpc}$, where $\Delta_{200}$ is the overdensity (calculated at $z=3$) in units of the `virial' value 200 and $\dot{N}_\textrm{ion,56}$ is the QSO production rate of hydrogen ionizing photons, in units of $\dot{N}_\textrm{ion} = 10^{56} \; \textrm{s}^{-1}$ (A typical value for QSOs in our faint
sample, using ionizing luminosities estimated by scaling the \citealt{Lusso+15} composite spectrum to the UV continua of our quasars). \footnote{A more precise calculation of the Str\"omgren radius would also include a weak dependence on the clumping factor $C$ as $C^{-1/3}$; this leaves our conclusions unchanged for any clumping factor smaller than or comparable to $C = 1000$. We have adopted the case B recombination coefficient evaluated at $T = 2 \times 10^4 \; \textrm{K}$ as in \cite{OsterbrockFerland}.} This is much larger than the physical radii (roughly, half the sizes reported in Table \ref{tab:nebulae} as the extent is more like the diameter) of our nebulae and it is therefore safe to assume that the illuminated gas is almost completely ionized (ionized fraction $x_\textrm{ion} \simeq 1$, neutral fraction $x_\textrm{HI} \ll 1$), implying that (density-bound) recombination is an important emission mechanism throughout the extent of our nebulae. 
Note that, even for our fainter QSOs, the estimated flux of ionizing photons is sufficiently large that the time required to ionize the entire CGM is bound by the speed of light, giving $1.6\times10^5$ yr for a distance of 50 kpc. This is shorter than the typical lifetime of quasars as estimated using the proximity effect and fluorescence, with values ranging from $10^6$ to $3\times10^7$ years (e.g. \citealt{Goncalves2008,  Trainor2013, Borisova_angle,Khrykin2019}), though these observations do not exclude some flickering of AGN activity on shorter timescales (see e.g. \citealp{Schawinski2015}, who presented evidence to support individual AGN phases as short as $10^5$ years). Note that the illumination of the quasar does not have to be continuous to keep the CGM highly ionized, as long as the total duration of activity is sufficient ($> 10^5$ years) over the last recombination time ($\simeq10^6$ years, e.g. \citealp{Pezzulli2019}). As we expect the nebulae to be very highly ionized we are in the density-bound case, implying in particular that the amount of \lya\ emission produced by recombinations does not depend \emph{directly} on the ionizing luminosity of the quasar, but only on the properties (mass, temperature, density distribution) of the illuminated gas (see Sect. \ref{subsec:PhysicalProperties} below for a discussion of possible \emph{indirect} links between these quantities). It remains to be discussed whether the other two mechanisms (scattering and collisional excitation) may give contributions comparable to, or even larger than, recombination in at least some regions of the nebulae. 

As discussed for instance in \cite{Pezzulli2019}, in a highly ionized medium scattering can dominate over recombination, but only in the extreme case that the CGM is optically thin to the Ly$\alpha$ radiation itself.\footnote{The basic reason why scattering is subdominant in the optically thick case is that the outer portions of the nebula do not \emph{directly} see the central source. Of course, depending on frequency diffusion, the outer regions may be able to reprocess photons that were already scattered once in the inner regions. This latter effect however causes only a redistribution and no increase of the overall luminosity.} Some authors (e.g. \citealp{MUSEUM}) have suggested that a relation between the luminosity density at the Ly$\alpha$ peak (erg s$^{-1}$ Hz$^{-1}$) of the QSO and the overall luminosity of the nebula (as we see in the top-right panel of Fig. \ref{fig:correlations}) could be an indication that scattering may be the dominant powering mechanism. However, one must also consider the fact that the neutral fraction of the CGM scales inversely with the ionizing flux of the QSO. Therefore, relatively bright QSOs would produce more \lya\ photons, but, on the other hand, would be surrounded by a smaller number of scattering targets, which should go in the opposite direction and in principle destroy a linear correlation. Under the assumption the nebulae are totally powered by scattering we have computed the optical depth of the nebula at the wavelength of peak nebular emission ($\tau$). This is calculated from the ratio of the peak flux of the nebula to the flux of the quasar at the same wavelength. For the fainter quasar sample $\tau$ ranges from 0.21 to 0.84. This range is somewhat small given that the sample spans a factor of 15.8 in UV luminosity. Using the Spearman's rank coefficient, there is no significant evidence ($\rm p=0.17$) for an anti-correlation between the inferred scattered fraction and the UV luminosity of the quasar among the faint sample. A model with a constant value of $\tau$ independent of UV luminosity provides lower $\chi^2$ statistic than the linear anti-correlation model, but the fit is not significantly better. Hence, our faint sample does not provide strong evidence of scattering as the dominant mechanism powering Giant \lya\ Nebulae, but our analysis is insufficient to reject the hypothesis at this time. 

Additionally, scattering does not necessarily explain the correspondence in \lya\ line profiles between nebulae and their quasars (see Sect. \ref{sec:spectra}). The scattered \lya\ line profile would be the product of the central spectrum that is incident on the nebula and the frequency-dependent scattering probability, which would depend on the kinematics of the CGM. It is therefore possible that the scattered profile would be narrower than the central narrow \lya\ components or broader, including some of the broad \lya\ line and quasar continuum. The central and scattered \lya\ line profiles would only be similar if the emitting region and CGM happened to have similar velocity dispersions. The scattering-dominated hypothesis also leaves open the origin of these narrow components, and why they seem to be more common with decreasing quasar UV luminosity.

We finally need to discuss collisional excitation. The relative contributions of collisional excitation and recombination to the observed Ly$\alpha$ emissivity can be written as
\begin{equation}\label{collratio1}
\frac{j_\textrm{coll}}{j_\textrm{rec}} = \frac{x_\textrm{HI}}{1 - x_\textrm{HI}}\frac{q^\textrm{eff}_{\textrm{Ly}\alpha}(T)}{\alpha^\textrm{eff}_{\textrm{Ly}\alpha}(T)} \;,
\end{equation}
where $x_\textrm{HI}$ is the neutral fraction, while $q^\textrm{eff}_{\textrm{Ly}\alpha}$ and $\alpha^\textrm{eff}_{\textrm{Ly}\alpha}$ are the effective (temperature-dependent) collisional excitation and recombination coefficients, respectively. The "effective" recombination (collisional excitation) coefficient describes recombination (collisional excitation) events giving rise to the emission of a \lya\ photon. In a highly ionized medium, we can write (e.g.\ \citealt{Meiksin2009}) $x_\textrm{HI} \simeq \alpha^\textrm{rec} n / \Gamma$, where $\alpha^\textrm{rec}$ is the total (case A) recombination coefficient, while $\Gamma = 10^{-9} \; r_{50}^{-2} \; s^{-1}$ is the photoionization rate associated to the typical ionizing luminosity of our faint QSOs, scaled to a fiducial distance of $50 \; \textrm{kpc}$ and $n$ is the total hydrogen number density. Equation \eqref{collratio1} therefore becomes

\begin{equation}\label{collratio2}
\frac{j_\textrm{coll}}{j_\textrm{rec}} = \frac{1}{\eta_{\textrm{Ly}\alpha}(T) }\frac{n q^\textrm{eff}_{\textrm{Ly}\alpha}(T)}{\Gamma} \;,
\end{equation}

\noindent where $\eta_{\textrm{Ly}\alpha}$ is the (temperature dependent) Ly$\alpha$ production probability per recombination, i.e. $\eta_{\textrm{Ly}\alpha} = \alpha^\textrm{eff}_{\textrm{Ly}\alpha}/\alpha^\textrm{rec}$. While $\eta_{\textrm{Ly}\alpha}(T)$ is a factor of order unity with a relatively small dependence on temperature, $q^\textrm{eff}_{\textrm{Ly}\alpha}(T)$ increases by orders of magnitude from $q^\textrm{eff}_{\textrm{Ly}\alpha}(T) = 2 \times 10^{-13} \; \textrm{cm}^3 \; \textrm{s}^{-1}$ at $T = 10^4 \; \textrm{K}$ to $q^\textrm{eff}_{\textrm{Ly}\alpha}(T) = 2 \times 10^{-9} \; \textrm{cm}^3 \; \textrm{s}^{-1}$ for $T = 5 \times 10^4 \; \textrm{K}$ (e.g.\ \citealt{Cantalupo+08}). For a typical density of the cold, pressure-confined CGM, $n = 0.1 \; \textrm{cm}^{-3}$ \citep{Cantalupo2019}, even assuming a rather high temperature, $T = 5 \times 10^4 \; \textrm{K}$, collisional excitation is at most comparable to recombination radiation at $r = 50 \; \textrm{kpc}$ and definitely negligible at any smaller distance or lower temperature. We plan to investigate these aspects in more detail with radiation-hydrodynamic simulations (Sarpas et al. in prep).

It is also possible that the dominant emission mechanism is different for the faint and bright samples, which could perhaps contribute to explaining the trend of increasing nebular SB with increasing QSO UV luminosity. A possible argument in favour of this option is the following. As mentioned above, within the illuminated (almost completely ionized) region, the small residual neutral fraction is expected to vary, to first order, in inverse proportion to the ionizing luminosity of the quasar. Depending on the temperature and, most importantly, on the detailed kinematics of the CGM, this might result in a transition in the optical depth to Ly$\alpha$ photons, from a large optical depth for faint QSOs to a relatively small optical depth for luminous ones. As shown in \cite{Pezzulli2019}, such a change could imply a transition from a recombination-dominated regime to a scattering-dominated regime. In the optically thin limit, scattering has $\sim 1 \; \textrm{dex}$ larger emissivity than recombination, which would be sufficient to explain the trend in Fig. \ref{fig:correlations} (top left). Interestingly, this scenario makes the very distinctive prediction that the $\textrm{H}\alpha/\textrm{Ly}\alpha$ ratio of the nebulae should decrease with increasing QSO luminosity, which makes it testable with future observations.

\subsection{Secondary parameters dependent on QSO luminosity}\label{subsec:PhysicalProperties}

As we discussed in Sect. \ref{sec:selection}, our sample was primarily selected on $i_{SDSS}$ magnitude. In interpreting our results, it is important to consider if the detected luminosity dependence of our nebulae is directly caused by the decreased quasar luminosity, or if the QSO luminosity correlates with some secondary property, which then drives the observed differences in the nebulae. In this section we consider the effects of the opening angle of illumination by the QSO and the halo masses of the host galaxies.

The illumination of quasars is believed to have a bi-conical geometry, due to a dusty obscuring torus encircling the AGN. The amount of light incident on the CGM is therefore related to the solid angle subtended by the quasar emission. If the opening angle is reduced, in a scenario where the nebular emission is dominated by scattering of \lya\ photons from the quasar, this could lead to a reduced nebula luminosity as there are fewer photons available to be scattered. A similar consideration also applies in the case of recombination, as the volume of the photoionized emitting region will be reduced. If the opening angle of quasars varies as a function of UV luminosity (or \lya\ peak luminosity) then trends between quasar luminosity and nebular properties are expected. For hyper-luminous quasars some limited constraints from fluorescent galaxies exist, indicating opening angles of $\theta>30^\circ$ \citep{Trainor2013,Borisova_angle}. In addition to trying to infer the opening angle of individual QSOs, one can determine the same quantity in a statistical way, by considering the fraction of the AGN population (at fixed intrinsic luminosity) that is obscured. It has been suggested that the obscured fraction of QSOs may be a decreasing function of their luminosity (\citealt{Ueda2003}; \citealt{Ichikawa+19}). However, these studies show that at even at the lowest luminosities probed in our samples the obscured fraction is $<0.5$ \citep{Ueda2014}. Therefore, it does not seem viable that a varying opening angle can explain the observed differences in nebular luminosities, which cover more than an order of magnitude. A luminosity-dependent opening angle may contribute to the observed relations, but alone it probably cannot fully account for them quantitatively. \footnote{If the opening angle of fainter quasars is smaller, one might expect to see a difference in nebular asymmetry with quasar luminosity. On the one hand, the decreased illuminated solid angle would restrict the range of inclinations where objects would still appear as unobscured (type-I) quasars, which should mean the faint quasars are viewed closer to their axes of symmetry. On the other hand, the shrinking ionization cones would decrease the overall isotropy of the ionized volume. The relative importance of these two opposing effects is unclear without modeling.  
As discussed in Appendix \ref{ap:asym} we see no significant evidence of a difference in asymmetry parameter, $\alpha$, between bright and fainter samples. }

Among the large-scale physical properties which may have an impact on the CGM and also may be directly or indirectly related to QSO luminosity, the most obvious is halo mass. Some theoretical models would predict a dependence of QSO host halo mass on QSO luminosity (e.g.\ \citealt{ConroyWhite2013}), although observations have revealed no significant evidence for luminosity dependence (e.g.\ \citealt{Chehade2016}; \citealt{Uchiyama+18}; \citealt{He+18}). The question regarding the mass of halos hosting the most luminous QSO at $z\simeq 3$ is particularly open, with different studies finding masses that are discrepant by as much as one order of magnitude (see e.g.\ \citealt{Pezzulli2019} for a discussion). If more luminous QSOs at $z\simeq3$ systematically live in more massive halos than fainter QSOs at the same redshift, then it is possible that the nebulae around them are more luminous because of the increased CGM mass. It is important to emphasize, however, that an increasing halo mass is by itself no guarantee of a higher nebular luminosity. Many important factors other than the total mass are expected to change at the same time, sometimes producing an opposite effect. One important example is the cold gas fraction, which many current models predict to decrease dramatically at sufficiently large masses (e.g. \citealt{BirnboimDekel2003}), which would eventually result in a drop of the Ly$\alpha$ luminosity. The possibility that the trends in Fig. \ref{fig:correlations} are being driven by halo mass is therefore intriguing, but definitely requires further investigation from both observational and theoretical perspectives.

\subsection{Correlations between nebula and QSO \texorpdfstring{Ly$\alpha$}{Lya} emission}\label{subsec:CentralEmission}

As described in Sect. \ref{subsec:lumdep}, we have found significant correlations between the quasar peak \lya\ luminosity density and the surface brightness of the nebulae measured between 20 and 50 pkpc. Perhaps the most striking result of Fig. \ref{fig:correlations} (right, middle panel) is that the scatter in this relation is smaller for the faint quasars, despite the lower signal-to-noise ratios of the measurements. In Sect. \ref{sec:spectra} it was noted that many of the faint quasars possess \lya\ lines with a narrow and a broad component. Fig. \ref{fig:spectra} also showed that in many cases (e.g. \#\csname J2305\endcsname, \csname J1307\endcsname\ and \csname J1000\endcsname\cmmnt{#1, 6 and 12}) the spectrum of the nebula seemed to follow the same line profile as the narrow component of the quasar line. 

We suggest that the narrow component of the \lya\ line, observed in the quasar spectrum, is due to the contribution from the nebula, which extends to smaller radii than we can examine due to our limited spatial resolution. Note that this inner emission could arise either in the host galaxy ISM or the inner CGM. This hypothesis would explain the correlation between SB$_{\rm outer}$ and $\rm L^{peak}_{Ly\alpha;QSO}$ as follows. For the faintest quasars the peak of the \lya\ would be dominated by the nebular component, and so these properties would be linearly correlated (assuming a fixed surface brightness profile and line width). As Fig. \ref{fig:correlations} (right middle row) shows, the fainter quasars roughly follow a linear relation, with some scatter beyond the observational uncertainties. As the UV luminosities of the quasars increase, $\rm L^{peak}_{Ly\alpha;QSO}$ will be boosted because the peak \lya\ luminosity density is the sum of the narrow and broad \lya\ lines and the underlying UV continuum. Hence, brighter quasars will move to the right of the linear relation. For the highest luminosities, the broad line should completely drown-out the narrow component, and so no correlation should persist. This is consistent with the bright \citetalias{Borisova} sample shown in the same figure, and the results of \cite{MUSEUM}. As a viability check we have included the exotic quasar of \cite{Marino2019}. This broad line quasar lacks a broad \lya\ line as it is totally absorbed by a pDLA. The remaining narrow \lya\ line is interpreted to be emission from the nebula, which can be traced all the way to the quasar position, due to the pDLA acting as a coronagraph. The green line in Fig. \ref{fig:correlations} (right middle row) shows the result of scaling the surface brightness and peak \lya\ of this quasar linearly. This extrapolation is a reasonably good description of the correlation for the fainter quasars, lending credit to the hypothesis that the nebulae are contributing to the peak of the quasars' \lya\ lines. Similar conclusions have been drawn for radio-loud QSOs \citep{Heckman1991}, based on the correspondence between the spectral profile of the narrow central component and extended emission. \cite{Fathivavsari2016} studied six QSOs with pDLAs, which block the broad \lya\ line of the AGN, and found that the remaining narrow central \lya\ was correlated linearly with the brightness of the nebula. \cite{Christensen2006} also detected tentative evidence of a correlation between QSO \lya\ luminosity and that of the extended \lya\ among radio quiet objects. In this study we have targeted intrinsically fainter QSOs allowing us to see this effect and observe both lines of evidence simultaneously, namely the correlated spectral profiles of nebulae and QSOs and the linear relation between the central \lya\ and the nebula luminosities. 

In \cite{Marino2019} it was observed that in the \lya\ surface brightness and kinematics there was no change in behaviour at small radii, suggesting that either the contribution from the ISM was negligible or there was a smooth transition from the CGM to the ISM. If the same behaviour extends to our faint sample and the narrow central emission is coming from the inner extent of the nebulae then we cannot determine if the emission arises from the ISM or CGM. While the CGM origin scenario naturally explains the linear correlation between $\rm L^{peak}_{Ly\alpha;QSO}$ and surface brightness, this correlation could also be understood if these narrow components come from the ISM of the host galaxies. In the early Universe, when accretion of gas through the CGM is believed to be a dominant process shaping the properties of galaxies, we expect the ISM mass to be correlated to the mass of the CGM. 

\subsection{Future Work}

In this paper we have uncovered connections between the properties of Giant \lya\ Nebulae and their host quasars. There are, however, a number of outstanding questions that we have been unable to answer conclusively. This initial study into the nebulae of faint quasars did not include any discussion of kinematics, as we have left this to future studies with deeper observations with high S/N maps. Another interesting avenue is to extend this study to even fainter objects; pushing to even lower ionizing luminosity may reveal a change in behaviour, when a quasar is no longer able to keep the CGM highly ionized. Further insight may emerge from deeper comparisons between quasar nebulae and the nebulae around type-II AGN. \cite{denBrok2020} have done an initial comparison of type I and II AGN with MUSE, however the Type I comparison sample (\citetalias{Borisova}) had much higher intrinsic luminosities than the Type II objects, due to the manner in which the samples were selected. A uniform sample of Type I and II AGN selected by intrinsic x-ray luminosity would allow for further tests of the AGN unification scheme. One of the most prominent issues surrounding quasar nebulae is the nature of the dominant mechanism powering the \lya\ emission in these nebulae,  which is believed to be either resonant scattering of photons from the quasar or recombination. From an observational perspective, the best remaining method of determining the dominant process is to search for non-resonant recombination lines, such as H$\rm \alpha$. For the $z>3$ MUSE nebulae this will only be possible in the JWST era, but for $z\simeq2$ nebulae (e.g. \citealp{Cai2019}) this can be done from the ground. Today it is possible to identify these lower redshift nebulae with KCWI, but the power of these studies would be enhanced with a more capable instrument such as BlueMUSE \citep{BlueMUSE}. It may also be possible to constrain the contribution of resonant scattering by studying the polarisation of the \lya\ emission, as was done for Lyman $\alpha$ Blobs \citep{Hayes2011}. 

\section{Conclusions}

We report the results of a MUSE survey of twelve $z\simeq3.15$ faint quasars, extending studies of giant \lya\ nebulae beyond the brightest quasars studied so far. Our sample was selected to have $3.1<z<3.25$, $20<i_{\rm SDSS}<23$ and a range of QSO \lya\ line strengths. We report the following results and conclusions:

\begin{itemize}
  \item We have detected \lya\ nebulae with 100\% success rate, even though we have targeted quasars at the limit of SDSS spectroscopically confirmed quasars. The detected nebulae are smaller in maximum detected extent than those around bright quasars, with a median size of 100 pkpc in comparison to 180 pkpc for the bright sample (Fig. \ref{fig:SB_maps}). This appears to be due to lower surface brightness combined with observational limits, rather than the nebulae appearing more truncated.\\
  
  \item The median surface brightness of the nebulae between 20 and 50 pkpc is only 5.4 times fainter than for the bright quasars, despite the median UV continuum being 3.7 magnitudes fainter (i.e. a factor of $\simeq$ 30 lower in luminosity). The shapes of the surface brightness profiles are consistent across the wide range in luminosity (Fig. \ref{fig:SB_comp}). \\
  
  \item The correlation between the brightness of the nebulae and the $i$ band luminosity of the QSOs (Fig. \ref{fig:correlations}) could be related to a luminosity dependence of the quasar halo mass, or is perhaps evidence of a transition from a recombination-dominated regime to a scattering-dominated regime with increasing ionization and thus mean free path of \lya\ photons. A luminosity dependence of the opening angle of quasars may contribute to the observed trend, but based on literature constraints it seems unlikely to be capable of explaining the magnitude of the change in nebular luminosity. \\
  
  \item We have found a significant relation between the peak flux of the quasar \lya\ line and the surface brightness of the nebulae (Fig. \ref{fig:correlations}). The scatter in this relation is smaller for the fainter quasar sample. We suggest that this could be explained by the unresolved inner parts of the nebulae contributing significantly to the narrow components of the observed central \lya\ lines, which are present in most of our quasars. This is strengthened by the visible correspondence in line profile between the nebulae and the narrow quasar component seen in some cases (Fig. \ref{fig:spectra}). This narrow central emission could be understood as emission from the host galaxy's ISM or from the inner regions of the CGM. The linear correlation could also be evidence for a model where the \lya\ emission is dominated by resonant scattering, however after accounting for the changing ionization parameter, the evidence for scattering is inconclusive. \\
  
\end{itemize}

\noindent We have successfully detected the dependence of the properties of Giant \lya\ Nebulae on the luminosity of the illuminating QSO. These observations place key constraints on the CGM of quasars at high redshift. The exact physics behind the luminosity dependencies is unclear, as there are multiple complicating factors which prevent us from drawing definitive conclusions and further observational work is needed to break degeneracies present in the modeling. Future studies of H$\alpha$ emission offer the most promising means to conclusively determine the dominant mechanism powering these nebulae and whether this changes for nebulae around QSOs of different luminosity. The ratio of \lya\ to H$\alpha$ will be lower in the recombination scenario than if scattering is dominant. In the extreme case where the nebulae are powered entirely by scattering, we would not expect to observe extended H$\alpha$ as the line is non-resonant. Once the degeneracy in the emission process is resolved, our faint quasar sample could be used to put novel constraints on the emission geometry of quasars across a wide range in luminosity and the physical properties of the CGM (mass, temperature and density distribution).

\section*{Acknowledgements}

The authors thank Daichi Kashino, for providing access to unpublished zCOSMOS Deep data, and Jakob S. den Brok for sharing code used in \cite{denBrok2020}. GP and SC acknowledge the support of the Swiss National Science Foundation [grant PP00P2163824]. SM is supported by the Experienced Researchers Fellowship, Alexander von Humboldt-Stiftung, Germany. This work is based on observations collected at the European Organisation for Astronomical Research  in the Southern Hemisphere under the MUSE GTO programme. The major analysis and production of figures in this work was conducted in Python, using standard libraries which include NumPy \citep{numpy}, SciPy \citep{SciPy}, Matplotlib \citep{Matplotlib} and the interactive command shell IPython \citep{ipython}. This research also made use of Astropy, a community-developed core Python package for Astronomy \citep{astropy}, and Photutils, an Astropy package for detection and photometry of astronomical sources \citep{photutils}. The python interface {\sc dustmaps} \citep{pydustmaps} was used to query galactic extinction maps. {\sc topcat}, a graphical tool for manipulating tabular data, was also utilized in this analysis \citep{TOPCAT}. This research has made use of the "Aladin sky atlas" developed at CDS, Strasbourg Observatory, France \citep{Aladin}.

\section*{Data availability}
The data underlying this article were retrieved from the ESO archive. The raw ESO data can be accessed through the ESO Science Archive Facility\footnote{\url{http://archive.eso.org/cms.html}} as part of program IDs 0101.A-0203, 0102.A-0448 and 0103.A-0272. The derived data generated in this research will be shared on reasonable request to the corresponding author.


\bibliographystyle{mnras}
\bibliography{bibtex}




\appendix{}

\newpage
\section{Remeasured data of bright quasars} \label{ap:A}
\input{tables/tab3_measurements_Bor}
Throughout this work we have made extensive comparisons to the sample of Giant \lya\ Nebulae around bright quasars of \citetalias{Borisova}. The quantities plotted in Figs. \ref{fig:SB_comp} and \ref{fig:correlations} have been remeasured from the original datacubes of \citetalias{Borisova}. For completeness we provide these values in Table \ref{tab:nebulae_borisova}. Note these values are corrected for Galactic extinction and were calculated using slightly different methods, so some values differ with respect to those presented in \citetalias{Borisova}. However, remeasuring these quantities allows us to ensure the comparisons with the faint sample are due to real differences and not to subtle differences in methodology. The sizes of the nebulae, defined by the maximum projected distance across the 3D mask, are typically slightly larger than reported in \citetalias{Borisova}. This is due to using less conservative masking when estimating their extents, consistent with the analysis of the faint quasar sample.  

\vspace{-0.4cm}
\section{Quantifying Nebular Asymmetry} \label{ap:asym}

\begin{figure}
	\centering
	\includegraphics[width=\columnwidth]{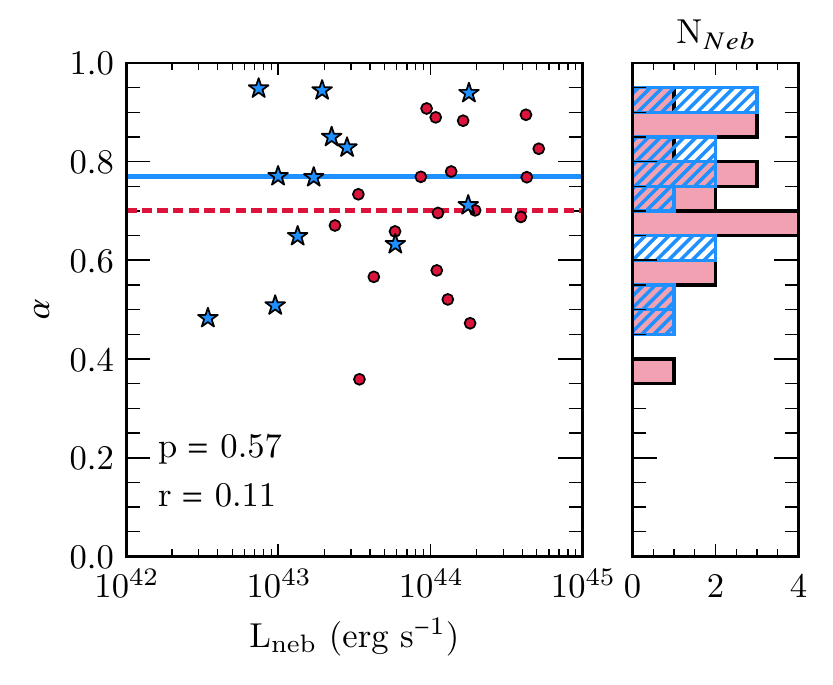}
	\vspace{-0.5cm}
	\caption{\textbf{(Left)} The nebula \lya\ luminosity against the nebula asymmetry parameter $\alpha$ for both the fainter sample (blue stars) and the bright sample (red circles, \protect\citetalias{Borisova}). The significance of the correlation is listed with the Spearman's correlation coefficient r, while the p-values indicate the significance at which the null hypotheses of no correlation can be rejected using this test. This example shows there is no significant evidence of a correlation between nebular luminosity and asymmetry. The horizontal lines show the median $\alpha$ of the faint (blue solid) and bright (red dashed) quasar samples. \textbf{(Right)} The $\alpha$ distributions of the faint (blue-hashed) and bright (red solid) quasar nebulae.}
	\label{fig:asym}
	\vspace{-1em}
\end{figure}

As noted in Sect. \ref{sec:morph}, the morphologies of the faint quasar nebulae appears to be very similar to those of the bright quasar sample (\citetalias{Borisova}), but with a lower surface brightness. In this section we attempt to quantify the asymmetry of these nebulae, using methods from the literature. \cite{MUSEUM} utilized a dimensionless asymmetry parameter, $\alpha$, which is calculated from the second order moments of the optimally extracted images. $\alpha$ is related to the ellipticity of the light distribution. We have applied a modified version of this procedure, described in \cite{denBrok2020}, which does not use flux weighting but instead uses the projected 3D mask (see Sect. \ref{sec:neb_detect}). Also, this method calculates the moments from the position of the quasar (or AGN), instead of the nebula centroid. We selected this variation as we were concerned that our faint nebulae would result in noisy centroids, thus injecting artificial scatter into the calculated values of $\alpha$. 

Fig. \ref{fig:asym} shows the calculated asymmetry parameters ($\alpha$) for the bright and fainter quasar nebulae, as a function of nebula luminosity, and the distribution of $\alpha$ for each sample. The medians of the two populations show that, on average, the fainter quasar nebulae are sightly more circularly symmetric. A Welch's t-test returns a p-value of $0.40$, indicating that there is no significant evidence to reject the null hypothesis that both samples have the same mean $\alpha$. A Kolmogorov-Smirnov (KS) test reveals that there is also no significant evidence of the $\alpha$ distributions being different between bright and faint quasar samples ($\rm p_{KS}=0.68$). We also repeated these tests using a flux-weighted $\alpha$ statistic (see \citealp{MUSEUM}), while this provided somewhat stronger evidence for differences between the samples ($\rm p_{t}=0.21$ and $\rm p_{KS}=0.22$), it is still not significant. We also assessed the evidence for correlations between nebula and quasar parameters and their $\alpha$ asymmetry values using a Spearman's rank correlation test (following Sect. \ref{subsec:lumdep}). The nebula \lya\ luminosities, the (redshift-corrected) surface brightness measured between 20 and 50 pkpc, the QSO $i$ band absolute magnitude and the peak luminosity density of the QSO \lya\ line were all compared to both the un-weighted and flux-weighted $\alpha$ values. We find no significant correlations, with p-values ranging between 0.41 and 0.98. 

We conclude from this investigation that there is no significant evidence for a difference in asymmetry between bright and fainter quasar nebulae. We note that $\alpha$ is not a general test of asymmetry, it is insensitive to non-circularity with three-fold symmetry or higher. Future work with higher SNR data may take a more comprehensive approach, such as the test presented in \cite{denBrok2020} using Fourier decomposition.

\bsp	
\label{lastpage}
\end{document}

%% file: tables/tab1_sample_MUSEphot.tex
\begin{table*}
\caption{Key properties of our sample of faint QSOs targeted with MUSE. $^{\rm a}$Taken from the SDSS database \protect\citep{SDSS_DR15}, with the exception of \#12, for which the data comes from zCOSMOS Deep. $^{\rm b}$Measured with aperture photometry from the MUSE datacubes with a 3'' radius aperture. $^{\rm c}$The peak specific flux of the QSO Ly$\alpha$ line, measured from the MUSE datacubes by extracting a QSO spectrum with a 2'' radius aperture. $^{\rm d}$The $i$ band absolute magnitude, K-corrected to $z=2$ following \protect\cite{Ross2013}. $^{\rm e}$ The Galactic extinction in the $i$ band using the \protect\cite{Planckdust} extinction map and the dust law from \protect\cite{CCM89}, with $R(V)=3.1$. $^{\rm f}$ Values corrected for Galactic extinction. $^{\rm g}$ The spatial FWHM measured in collapsed 4800-5300 \AA\ images made from the MUSE datacubes, this band is centered on the median quasar Ly$\alpha$ wavelength. }
 \begin{tabular}{cccccccccc} 
\hline
Id & Name & RA$^{\rm a}$ & Dec$^{\rm a}$ & $z$$^{\rm a}$ & $i$$^{\rm b,f}$ & L$\rm ^{peak}_{Ly\alpha;QSO}$$^{\rm c,f}$ & M$_i (z=2)$$^{\rm d,f}$ & A$_i$$^{\rm e}$ & Seeing FWHM$^{\rm g}$ \\
  \# &   &   &   &  & (mag) & (erg s$^{-1}$ Hz$^{-1}$) &  & (mag) & (arcsec) \\
\hline
  1 & J0827$+$0716 & 08:27:10.97 & $+$07:16:50.0 & 3.146 & 19.69 & 2.88$\rm \times 10^{+31}$ & -27.13 & 0.09 & 1.38 \\ 
  2 & J0859$-$0018 & 08:59:36.77 & $-$00:18:57.2 & 3.187 & 19.74 & 1.25$\rm \times 10^{+32}$ & -27.11 & 0.08 & 1.07 \\ 
  3 & J0002$-$0721 & 00:02:32.50 & $-$07:21:20.4 & 3.152 & 20.34 & 2.04$\rm \times 10^{+31}$ & -26.48 & 0.09 & 1.13 \\ 
  4 & J0759$+$0605 & 07:59:49.98 & $+$06:05:47.3 & 3.233 & 20.71 & 7.75$\rm \times 10^{+30}$ & -26.18 & 0.05 & 1.12 \\ 
  5 & J0854$+$0328 & 08:54:38.63 & $+$03:28:14.9 & 3.179 & 20.95 & 1.36$\rm \times 10^{+31}$ & -25.89 & 0.10 & 0.80 \\ 
  6 & J1307$+$0202 & 13:07:01.71 & $+$02:02:41.1 & 3.230 & 21.17 & 3.71$\rm \times 10^{+31}$ & -25.71 & 0.06 & 0.80 \\ 
  7 & J1352$-$0110 & 13:52:57.97 & $-$01:10:40.1 & 3.144 & 21.47 & 7.19$\rm \times 10^{+30}$ & -25.34 & 0.11 & 0.99 \\ 
  8 & J2305$-$0034 & 23:05:06.76 & $-$00:34:54.9 & 3.196 & 21.72 & 2.44$\rm \times 10^{+31}$ & -25.13 & 0.11 & 1.03 \\ 
  9 & J0148$-$0055 & 01:48:08.97 & $-$00:55:08.5 & 3.155 & 21.74 & 2.40$\rm \times 10^{+31}$ & -25.09 & 0.07 & 0.96 \\ 
  10 & J0826$+$0005 & 08:26:00.97 & $+$00:05:08.6 & 3.115 & 21.90 & 8.55$\rm \times 10^{+30}$ & -24.88 & 0.10 & 0.74 \\ 
  11 & J0923$+$0011 & 09:23:00.35 & $+$00:11:56.1 & 3.151 & 22.14 & 1.19$\rm \times 10^{+31}$ & -24.67 & 0.06 & 1.15 \\ 
  12 & J1000$+$0223 & 10:00:50.59 & $+$02:23:29.0 & 3.095 & 22.99 & 3.58$\rm \times 10^{+30}$ & -23.78 & 0.04 & 0.88 \\ 
  \hline
 \end{tabular}
 \label{tab:sample}
\end{table*}

%% file: tables/tab2_measurements_MUSEphot.tex
\begin{table*}
\caption{Measured properties of the \lya\ nebulae. 
$^{\rm a}$ The total \lya\ luminosity of the nebula, taken by integrating the flux in the continuum-subtracted cube within the 3D mask. 
$^{\rm b}$ The maximum linear extent of each nebula. This is the largest projected distance between pixels within the 3D mask. $^{\rm c}$ The mean \lya\ surface brightness in two annuli, between 10 and 15 pkpc and 20 and 50 pkpc. These values are corrected for redshift dimming to $z=3$. The upper limit for J0826+0005 corresponds to 3$\sigma$. $^{\rm d}$ Corrected for Galactic extinction, using the \protect\cite{Planckdust} dust map and the extinction law of \protect\cite{CCM89}, calculated at the flux weighted wavelength of each nebula with $R(V)=3.1$. }
 \begin{tabular}{cccrcc} 
\hline
Id & Name & L$_{\rm Ly\alpha,neb}$$^{\rm a,d}$ & Size$^{\rm b}$ & SB(10-15 pkpc)$[(1+z)/4]^4$  $^{\rm c,d}$ & SB(20-50 pkpc)$[(1+z)/4]^4$  $^{\rm c,d}$ \\
 \# &   & (erg s$^{-1}$) & (pkpc) & (erg s$^{-1}$ cm$^{-2}$ arcsec$^{-2}$) & (erg s$^{-1}$ cm$^{-2}$ arcsec$^{-2}$) \\
\hline
  1 & J0827$+$0716 & 2.24$\rm \times 10^{+43}$ & 130 & 1.06$\pm$0.06$\rm \times 10^{-17}$ & 2.69$\pm$0.12$\rm \times 10^{-18}$ \\ 
  2 & J0859$-$0018 & 1.77$\rm \times 10^{+44}$ & 140 & 6.90$\pm$0.08$\rm \times 10^{-17}$ & 7.43$\pm$0.18$\rm \times 10^{-18}$ \\ 
  3 & J0002$-$0721 & 1.94$\rm \times 10^{+43}$ & 90 & 9.03$\pm$0.52$\rm \times 10^{-18}$ & 1.35$\pm$0.12$\rm \times 10^{-18}$ \\ 
  4 & J0759$+$0605 & 3.44$\rm \times 10^{+42}$ & 60 & 2.45$\pm$0.31$\rm \times 10^{-18}$ & 3.68$\pm$0.71$\rm \times 10^{-19}$ \\ 
  5 & J0854$+$0328 & 9.54$\rm \times 10^{+42}$ & 150 & 4.95$\pm$0.40$\rm \times 10^{-18}$ & 1.41$\pm$0.09$\rm \times 10^{-18}$ \\ 
  6 & J1307$+$0202 & 1.79$\rm \times 10^{+44}$ & 190 & 4.93$\pm$0.07$\rm \times 10^{-17}$ & 1.08$\pm$0.02$\rm \times 10^{-17}$ \\ 
  7 & J1352$-$0110 & 7.42$\rm \times 10^{+42}$ & 70 & 3.20$\pm$0.42$\rm \times 10^{-18}$ & 3.54$\pm$0.90$\rm \times 10^{-19}$ \\ 
  8 & J2305$-$0034 & 2.83$\rm \times 10^{+43}$ & 110 & 3.21$\pm$0.06$\rm \times 10^{-17}$ & 2.02$\pm$0.15$\rm \times 10^{-18}$ \\ 
  9 & J0148$-$0055 & 5.87$\rm \times 10^{+43}$ & 130 & 3.12$\pm$0.06$\rm \times 10^{-17}$ & 1.36$\pm$0.14$\rm \times 10^{-18}$ \\ 
  10 & J0826$+$0005 & 9.96$\rm \times 10^{+42}$ & 60 & 6.18$\pm$0.50$\rm \times 10^{-18}$ & $<$3.51$\rm \times 10^{-19}$ \\ 
  11 & J0923$+$0011 & 1.71$\rm \times 10^{+43}$ & 90 & 8.87$\pm$0.38$\rm \times 10^{-18}$ & 4.57$\pm$0.88$\rm \times 10^{-19}$ \\ 
  12 & J1000$+$0223 & 1.34$\rm \times 10^{+43}$ & 90 & 3.38$\pm$0.36$\rm \times 10^{-18}$ & 5.28$\pm$0.86$\rm \times 10^{-19}$ \\ 
  \hline
 \end{tabular}
\label{tab:nebulae}
\end{table*}

%% file: tables/tab3_measurements_Bor.tex
\begin{table*}
\caption{Measured properties of the \lya\ nebulae of the bright quasar sample from \protect\citetalias{Borisova}. 
$^{\rm a}$The peak specific flux of the QSO Ly$\alpha$ line, measured from the MUSE datacubes by extracting a QSO spectrum with a 2'' radius aperture. 
$^{\rm b}$The $i$ band absolute magnitude, K-corrected to $z=2$ following \protect\cite{Ross2013}. 
$^{\rm c}$ The total \lya\ luminosity of the nebula, taken by integrating the flux in the continuum-subtracted cube within the 3D mask. 
$^{\rm d}$ The maximum linear extent of each nebula. This is the largest projected distance between pixels within the 3D mask. 
$^{\rm e}$ The mean \lya\ surface brightness in two annuli, between 10 and 15 pkpc and 20 and 50 pkpc. These values are corrected for redshift dimming. 
$^{\rm f}$ Corrected for Galactic extinction, using the \protect\citealp{Planckdust} dust map and the extinction law of \protect\citealp{CCM89}, calculated at the flux weighted wavelength of each nebula with $R(V)=3.1$.}
 \begin{tabular}{cccccrcc} 
\hline
Id & Name  & L$\rm ^{peak}_{Ly\alpha;QSO}$$^{\rm a,f}$ & M$_i (z=2)$$^{\rm b,f}$ & L$_{\rm Ly\alpha,neb}$$^{\rm c,f}$ & Size$^{\rm d}$ & SB(10-15 pkpc)$[(1+z)/4]^4$ $^{\rm e,f}$ & SB(20-50 pkpc)$[(1+z)/4]^4$ $^{\rm e,f}$ \\
 \# &    & (erg s$^{-1}$ Hz$^{-1}$) &   & (erg s$^{-1}$) & (pkpc) & (erg s$^{-1}$ cm$^{-2}$ arcsec$^{-2}$) & (erg s$^{-1}$ cm$^{-2}$ arcsec$^{-2}$) \\
\hline
  1 & CTS G18 & 3.62$\rm \times 10^{+32}$ & -30.27 & 1.97$\rm \times 10^{+44}$ & 260 & 4.40$\pm$0.14$\rm \times 10^{-17}$ & 6.70$\pm$0.14$\rm \times 10^{-18}$ \\ 
  2 & Q0041$-$2638 & 7.87$\rm \times 10^{+31}$ & -28.46 & 3.41$\rm \times 10^{+43}$ & 180 & 7.48$\pm$0.63$\rm \times 10^{-18}$ & 2.26$\pm$0.10$\rm \times 10^{-18}$ \\ 
  3 & Q0042$-$2627 & 8.47$\rm \times 10^{+31}$ & -28.27 & 1.82$\rm \times 10^{+44}$ & 330 & 3.42$\pm$0.08$\rm \times 10^{-17}$ & 7.34$\pm$0.14$\rm \times 10^{-18}$ \\ 
  4 & Q0055$-$269 & 2.26$\rm \times 10^{+32}$ & -29.59 & 3.93$\rm \times 10^{+44}$ & 190 & 1.03$\pm$0.02$\rm \times 10^{-16}$ & 1.70$\pm$0.02$\rm \times 10^{-17}$ \\ 
  5 & UM669 & 2.13$\rm \times 10^{+32}$ & -28.97 & 1.08$\rm \times 10^{+44}$ & 160 & 3.13$\pm$0.13$\rm \times 10^{-17}$ & 7.21$\pm$0.14$\rm \times 10^{-18}$ \\ 
  6 & J0124 & 1.59$\rm \times 10^{+32}$ & -29.06 & 4.29$\rm \times 10^{+44}$ & 160 & 1.08$\pm$0.02$\rm \times 10^{-16}$ & 3.07$\pm$0.04$\rm \times 10^{-17}$ \\ 
  7 & UM678 & 2.72$\rm \times 10^{+32}$ & -28.90 & 8.64$\rm \times 10^{+43}$ & 190 & 1.23$\pm$0.09$\rm \times 10^{-17}$ & 6.49$\pm$0.14$\rm \times 10^{-18}$ \\ 
  8 & CTS B27 & 1.92$\rm \times 10^{+32}$ & -28.88 & 9.43$\rm \times 10^{+43}$ & 160 & 2.36$\pm$0.06$\rm \times 10^{-17}$ & 5.64$\pm$0.11$\rm \times 10^{-18}$ \\ 
  9 & CTS A31 & 2.03$\rm \times 10^{+32}$ & -28.99 & 1.12$\rm \times 10^{+44}$ & 190 & 2.81$\pm$0.08$\rm \times 10^{-17}$ & 7.69$\pm$0.12$\rm \times 10^{-18}$ \\ 
  10 & CT 656 & 1.63$\rm \times 10^{+32}$ & -29.37 & 3.36$\rm \times 10^{+43}$ & 140 & 9.91$\pm$0.67$\rm \times 10^{-18}$ & 3.01$\pm$0.11$\rm \times 10^{-18}$ \\ 
  11 & ALW 11 & 1.03$\rm \times 10^{+32}$ & -28.84 & 5.85$\rm \times 10^{+43}$ & 130 & 2.35$\pm$0.07$\rm \times 10^{-17}$ & 3.00$\pm$0.13$\rm \times 10^{-18}$ \\ 
  12 & HE0940$-$1050 & 3.01$\rm \times 10^{+32}$ & -30.23 & 1.64$\rm \times 10^{+44}$ & 180 & 5.00$\pm$0.13$\rm \times 10^{-17}$ & 7.41$\pm$0.17$\rm \times 10^{-18}$ \\ 
  13 & BRI1108$-$07 & 1.81$\rm \times 10^{+32}$ & -28.97 & 1.30$\rm \times 10^{+44}$ & 170 & 4.11$\pm$0.20$\rm \times 10^{-17}$ & 8.58$\pm$0.36$\rm \times 10^{-18}$ \\ 
  14 & CTS R07 & 6.36$\rm \times 10^{+32}$ & -29.54 & 4.24$\rm \times 10^{+44}$ & 190 & 1.27$\pm$0.02$\rm \times 10^{-16}$ & 2.41$\pm$0.02$\rm \times 10^{-17}$ \\ 
  15 & Q1317$-$0507 & 2.85$\rm \times 10^{+32}$ & -29.60 & 4.24$\rm \times 10^{+43}$ & 150 & 1.26$\pm$0.12$\rm \times 10^{-17}$ & 3.49$\pm$0.20$\rm \times 10^{-18}$ \\ 
  16 & Q1621$-$0042 & 8.49$\rm \times 10^{+32}$ & -30.14 & 1.10$\rm \times 10^{+44}$ & 180 & 1.73$\pm$0.30$\rm \times 10^{-17}$ & 7.79$\pm$0.46$\rm \times 10^{-18}$ \\ 
  17 & CTS A11 & 2.30$\rm \times 10^{+32}$ & -29.03 & 2.36$\rm \times 10^{+43}$ & 170 & 8.56$\pm$0.77$\rm \times 10^{-18}$ & 2.19$\pm$0.10$\rm \times 10^{-18}$ \\ 
  R1 & PKS1937$-$101 & 9.50$\rm \times 10^{+32}$ & -30.73 & 5.15$\rm \times 10^{+44}$ & 160 & 1.28$\pm$0.03$\rm \times 10^{-16}$ & 3.10$\pm$0.05$\rm \times 10^{-17}$ \\ 
  R2 & QB2000$-$330 & 3.94$\rm \times 10^{+32}$ & -30.00 & 1.37$\rm \times 10^{+44}$ & 170 & 4.41$\pm$0.25$\rm \times 10^{-17}$ & 1.15$\pm$0.03$\rm \times 10^{-17}$ \\ 
  \hline
 \end{tabular}
 \label{tab:nebulae_borisova}
\end{table*}